\documentclass{emulateapj}

\newcommand{\ba}{\begin{eqnarray}}
\newcommand{\ea}{\end{eqnarray}}
\newcommand{\beq}{\begin{equation}}
\newcommand{\eeq}{\end{equation}}
\newcommand{\g}{\gamma}
\newcommand{\gp}{{\gamma^\prime}}
\newcommand{\gpp}{{\gamma_p^\prime}}
\newcommand{\dD}{{\delta_{\rm D}}}
\newcommand{\tp}{t^\prime}

\def\e{\epsilon}
\def\ep{\epsilon^\prime}

\begin{document}

\title{ Variable Gamma-ray Emission Induced by Ultra-High Energy Neutral Beams: Application to 4C +21.35}

\author{Charles D. Dermer\altaffilmark{1}, Kohta Murase\altaffilmark{2}, Hajime Takami\altaffilmark{3}}

\altaffiltext{1}{Code 7653, Space Science Division, U.S. Naval Research Laboratory, Washington, DC 20375, USA. e-mail: charles.dermer@nrl.navy.mil}
\altaffiltext{2}{Department of Physics, Center for Cosmology and Astro-Particle Physics, The Ohio State University, Columbus, OH 43210, USA. email: murase.2@osu.edu}
\altaffiltext{3}{Max Planck Institute for Physics, F\"ohringer Ring 6, 80805 Munich, Germany. email:takami@mpp.mpg.de. Affiliation as of 1 April 2012: Cosmophysics group, Theory Center, Institute of Particle and Nuclear Studies, KEK (High Energy Accelerator Research Organization), 1-1, Oho, Tsukuba 305-0801, Japan}

\begin{abstract}
The flat spectrum radio quasar (FSRQ) 4C +21.35 (PKS 1222+216) displays prominent nuclear infrared emission from $\approx 1200$ K dust. A 70 -- 400 GeV flare with $\approx 10$ min variations during half an hour of observations was found by the MAGIC telescopes, and GeV variability was observed on sub-day timescales with the Large Area Telescope on {\it Fermi}. We examine 4C +21.35, assuming that it is a source of ultra-high energy cosmic rays (UHECRs). UHECR proton acceleration in the inner jet powers a neutral beam of neutrinos, neutrons and $\gamma$ rays from $p\gamma$ photopion production. The radiative efficiency and production spectra of neutrals formed through photohadronic processes with isotropic external target photons of the broad line region and torus are calculated.  Secondary radiations made by this process have a beaming factor $\propto \delta_{\rm D}^5$, where $\delta_{\rm D}$ is the Doppler factor. The pair-production optical depth for $\g$ rays and the photopion efficiency for UHECR neutrons as they pass through external isotropic radiation fields are calculated. If target photons come from the broad line region and dust torus, large Doppler factors, $\dD\gtrsim 100$ are required to produce rapidly variable secondary radiation with isotropic luminosity $\gtrsim 10^{47}$ erg s$^{-1}$ at the pc scale.  The $\gamma$-ray spectra from leptonic secondaries are calculated from cascades initiated by the UHECR neutron beam at the pc-scale region and fit to the flaring spectrum of 4C +21.35. Detection of $\gtrsim 100$ TeV neutrinos from 4C +21.35 or other VHE blazars with IceCube or KM3NeT would confirm this scenario.
\end{abstract}


\section{Introduction} \label{sec:introduction}

Three flat spectrum radio quasars, 3C 279 \citep{alb08,ale11a}, PKS 1510-089 \citep{wag10,cor12}, 
and PKS 1222+216 \citep{ale11b}, have been detected in the very-high energy (VHE;  $\gtrsim 100$ GeV) regime. This last object, a blazar at redshift $z = 0.432$ also known as 4C +21.35, is remarkable for day-scale GeV flares monitored with the Large Area Telescope (LAT) on the {\it Fermi Gamma-ray Space Telescope}  in 2010 April and June \citep{tan11} that coincided with a period of high infrared activity \citep{car10}. Moreover, the {\it Major Atmospheric Gamma-ray Imaging Cherenkov (MAGIC)} telescopes observed a VHE flaring episode during a period of observation lasting about one-half hour on 2010 June 17 that showed $\approx$ factor-of-two variations on a 10 minute time scale \citep{ale11b}. The rapidly variable $\gamma$ rays reached large apparent isotropic $\gamma$-ray luminosities $L_\gamma \gtrsim 10^{47}$ erg s$^{-1}$ in the VHE band and $ \gtrsim 10^{48}$ erg s$^{-1}$ above 100 MeV.

The intense and highly variable $\gamma$-ray fluxes measured from blazars, including 4C +21.35, demonstrate that they are powerful accelerators of energetic particles. Energetic leptons make intense fluxes of synchrotron emission, with associated $\gamma$ rays formed by Compton processes 
from the same electrons.  As shown by {\it Fermi} LAT observations \citep{2011ApJ...743..171A}, 
the two-year time-averaged values of $L_\gamma$ of dozens of FSRQs exceed $ 10^{48}$ erg s$^{-1}$. During flaring episodes,  values of
 $L_\gamma \gtrsim 10^{49}$ erg s$^{-1}$ are not unusual \citep[the record-setting value is
$L_\gamma \approx 2\times 10^{50}$ erg s$^{-1}$ from the November 2010 outburst of 3C 454.3; ][]{2011ApJ...733L..26A}.
In addition to relativistic leptons, hadrons are also likely to be accelerated in blazar jets and make secondary $\gamma$ radiations by 
interacting with target photons of the internal and surrounding radiation fields. 
This emission will contribute to the formation of the 
spectral energy distribution (SED) 
\citep[see][for a recent review]{2010arXiv1006.5048B}, and the accelerated hadrons or secondary neutrons   
could escape from the acceleration region to become ultra-high energy cosmic rays (UHECRs) with energies $\gtrsim 10^{18}$ eV.

Typical quasars have broad optical and ultraviolet emission lines that originate from the broad line region (BLR) near the supermassive black hole. The rapid variability, as short as a few hours for 4C +21.35 as found in the {\it Fermi}-LAT data \citep{2011A&A...530A..77F,2011arXiv1110.4471F}, indicates that the high-energy $\g$ rays are formed close to the black hole within the BLR. But if this is the case, such high-energy photons would be strongly attenuated through the $\g\g\rightarrow e^+e^-$ process against scattered line and accretion-disk radiation at the sub-pc scale. We argue that the contradictory behaviors of rapid variability, large luminosity, and a production region distant from the central nucleus can be reconciled if the inner jet accelerates UHECRs that escape as neutrons, which then undergo photo-hadronic breakup 
in the infrared (IR) field of the dusty torus 
to make rapidly variable synchrotron $\g$ rays at GeV -- TeV energies on the pc scale. Furthermore, UHE 
$\gamma$ rays made in the inner jet can contribute to the cascade at the pc scale.

The recent observations of 4C +21.35 are summarized in Section \ref{sec:obs4C+2135}. 
An overview of the model, with estimates of the radiative efficiency and secondary photohadronic production spectra, is given in Section \ref{sec:model}. A detailed treatment of secondary production of neutrals is presented in Section \ref{sec:spec}, giving beaming factors and efficiencies of secondaries made in relativistic jets. Production spectra of neutrons and neutrinos by accelerated UHECRs 
are calculated. In Section \ref{sec:opacity}, formulas for and calculations of pair-production opacity and photohadronic efficiency of $\gamma$ rays and neutrons as they travel through the BLR are given. We show at the end of this Section that the BLR is highly opaque to GeV and TeV $\gamma$ rays, while TeV photons are still strongly attenuated at the pc scale. Ultra-high energy protons and neutrons escaping from the inner jet are shown to lose significant energy through photopion production at the pc scale.  In Section \ref{sec:sync}, the parsec-scale magnetic field values are derived that allow for sub-10 minute variability of lepton synchrotron radiation from electrons and positrons (hereafter referred to as ``electrons") formed as secondaries of photohadronic neutron production. Numerical fits to the MAGIC data of 4C +21.35 are presented. Discussion and summary are given in Section \ref{sec:discussion}.

We make the simplifying assumption throughout that the radiation fields are locally isotropic. This assumption becomes increasingly poor if the radiation energy density is dominated by anisotropically distributed photons impinging from behind. Thus, provided the emitting region is within the BLR radius ($R_{\rm BLR}$) for processes involving atomic-line or reprocessed accretion-disk radiation, or within the pc scale for processes involving the IR radiation field of the dusty torus, these results give an accurate determination of the $\g\g$ opacity and photopion efficiency. Note also, that for 4C +21.35 at $z = 0.432$, the luminosity distance $d_L = 2335$ Mpc $= 7.20\times 10^{27}$ cm  in a $\Lambda$CDM cosmology, where the Hubble constant $h = 0.72$, in units of 100 km s$^{-1}$ Mpc$^{-1}$, $\Omega_m = 0.27$ is the ratio of matter density to the critical density, and $\Omega_{\Lambda} = 0.73 = 1- \Omega_m$ is the ratio of cosmological constant to the critical density.

\section{Observations of 4C +21.35} \label{sec:obs4C+2135}

In this section,  the relevant observations and radiation fields in the nuclear region
of 4C +21.35 are described. 

\subsection{Observations}

The discovery of VHE emission between $\approx 70$ and 400 GeV with the MAGIC telescopes
\citep{mar10,ale11b} was anticipated by the report of VHE photons found in {\it Fermi} LAT data during a high state of 4C +21.35 in April 2010 \citep{nsv10}. The source was targeted with the MAGIC telescopes from 3 May 2010. The spectrum of the major VHE flare observed with MAGIC on 17 June 2010, which observed for only $\sim 30$ minutes, is well described by a single power law with photon number index $\alpha_\Gamma \cong 3.75$ that is hardened by about one unit when corrected for attenuation with a low-intensity model \citep{kd10} of the extragalactic  background light (EBL) at infrared and optical frequencies. The remarkably short 10 minute variability \cite[in comparison, 3C 279 displayed weak VHE variability on time scales of days; see][]{alb08} points to a production site near the central supermassive black hole where collimation by extreme processes can occur \citep{tav10}. Variability associated with the dynamical timescale of a $10^9 M_9$ Solar mass black hole is
$t_{\rm dyn} = (1+z) r_{\rm sh} / c = (1+z) 10^4 M_9$ s, where $r_{\rm sh}$ is the Schwarzschild radius of the black hole,  whereas variability on time scales shorter than $t_{\rm dyn}$ by factors of 10 or more has been observed in the rapidly variable BL Lac objects Mrk 501 \citep{aha07}, PKS 2155-304 \citep{aha07}, and Mrk 421 \citep{fos08}. For 4C +21.35, with a black hole mass estimated at $\approx 1.5\times 10^8 M_\odot$ \citep{2004ApJ...615L...9W}, the corresponding timescale is $\approx 2000$ s, so the VHE measurements indicate a marginally hyper-variable source.  

{  A higher black-hole mass is obtained from recent optical spectroscopy of J1224+2122 by
\citet{2012ApJ...748...49S}, using scaling relations derived from reverberation mapping of radio-quiet AGNs. 
From analysis of the FWHM of the H$\beta$ line, they deduce a black-hole mass for 4C +21.35 of 
$7.8(\pm 2.3) \times 10^8 M_\odot$, which does not include uncertainties in applying scaling relations to blazars. 
This value is considerably higher than the mass reported by  \citet{2004ApJ...615L...9W}, and would relax the energetics
based on the Eddington luminosity, but make the short timescale variability more problematic. Here we use the 
lower mass, and return to this point in Section 7.
}

The BLR radius $R_{\rm BLR}$ can be written in terms of the accretion-disk luminosity  $L_{\rm disk}$ as \citep{gt08}
\begin{equation}
R_{\rm BLR} \cong 10^{17} \sqrt{ {L_{\rm disk}\over 10^{45}{\rm ~erg~s}^{-1}}}\;{\rm cm}\;.
\label{RBLR}  
\end{equation}
Based on the H$\beta$ luminosity, $L_{\rm H\beta} \cong 2 \times 10^{43}$ erg s$^{-1}$, \citet{tan11} derive from the scaling relations of \citet{2004ApJ...615L...9W} and \cite{2006ApJ...646....8F} that the BLR luminosity $L_{\rm BLR} \approx 5 \times 10^{44}$ erg s$^{-1}$, and that the accretion disk luminosity $L_{\rm disk}$ is an order of magnitude greater, $L_{\rm disk} \approx 5\times 10^{45}$ erg s$^{-1}$. In comparison, \citet{tav11} estimate that $L_{\rm disk} \cong 5 \times 10^{46}$ erg s$^{-1}$, assuming the Swift UVOT spectrum from 4C +21.35 is the optically thick accretion-disk radiation field, in which case $R_{\rm BLR} \approx 7 \times 10^{17}$ cm $\approx 0.2$ pc. This represents super-Eddington luminosities, given that $L_{\rm Edd} \approx 2 \times 10^{46}$ erg s$^{-1}$ { unless the black-hole mass is $\gtrsim 4\times 10^8 M_\odot$.}  We use the smaller disk luminosity here, noting that a larger $L_{\rm disk}$ and BLR radius will improve photohadronic efficiency. 

More precise relations depend on specific lines and assumptions about the geometry of the BLR \citep[see, e.g., the recent discussion by][]{fos11}. In particular, we could use the rest-frame 5100 \AA$~$or 1350 \AA$~$continuum, or the high-ionization Lyman $\alpha$ line, as seen strongly in, e.g., 3C 454.3 \citep{bon11}. For definiteness, we focus on the Ly $\alpha$ line to illustrate BLR line effects, which typically carries $\gtrsim 20$\% of the BLR luminosity \citep{fra91}. Approximating the Ly $\alpha$/BLR line luminosity $L_{{\rm Ly}\alpha} \cong 0.1 L_{\rm disk}$. and defining 
\begin{equation}
\phi \;\equiv\; { L_{{\rm Ly}\alpha} /L_{\rm disk}\over (R_{{\rm Ly}\alpha} /R_{\rm BLR})^2}\;
\equiv \; 0.1\phi_{-1}
\label{LLyalpha}
\end{equation}
gives a Ly $\alpha$  photon field with energy density $u_{{\rm Ly}\alpha}\approx L_{{\rm Ly}\alpha} /4\pi R_{{\rm Ly}\alpha}^2c \cong 0.026\phi_{-1}$  erg cm$^{-3}$.

The magnetic field $B^\prime$ in the comoving jet frame can be determined if the GeV $\gamma$ rays are Compton-scattered external radiation and the lower energy optical radiation is  synchrotron radiation made by the same nonthermal electrons making the $\gamma$ rays at GeV energies. The Compton dominance parameter $A_{\rm C}$ is defined as the ratio of the $\gamma$-ray luminosity (assumed to originate from external Compton scattering) to the radio/X-ray synchrotron luminosity. For 4C +21.35, $A_{\rm C}$ reaches values as large as $\approx 100$ \citep[see multiwavelength SED using contemporaneous data in][]{tav11}.
If the external radiation field scattered by these electrons are the Ly $\alpha$ photons, then $A_{\rm C} \approx u^\prime_{{\rm Ly}\alpha}/u_B^\prime$, with  $u^\prime_{{\rm Ly}\alpha}\approx 4\Gamma^2 u_{{\rm Ly}\alpha}/3$,
and $u^\prime_B = B^{\prime 2}/8\pi$ \citep[e.g.,][]{2009ApJ...704...38S}. Thus 
the comoving magnetic field for a blazar jet with bulk Lorentz factor $\Gamma$ is
\begin{equation}
B^\prime \cong 9.3 \left({\Gamma\over100} \right) \;\sqrt{\phi_{-1}\over (A_{\rm C}/100)}\;\;{\rm G}\;.
\label{Bprime}
\end{equation} 

Modeling 5 -- 35$\,\mu$ Spitzer, SDSS, 2MASS, and Swift UVOT data of 4C +21.35, \cite{mal11} decompose its spectrum into a nonthermal power-law and two-temperature dust model. Hot dust with $T\approx 1200$ K radiates $\approx 8\times 10^{45}$ erg s$^{-1}$ from a pc-scale region, and a second warm dust component radiates  $\approx 10^{45}$ erg s$^{-1}$ at $T\approx 660$ K on the same scale. { The IR component, which has been proposed as a target in external Compton scenarios \citep{2000ApJ...545..107B, 2009ApJ...704...38S}, is especially important for UHECR scenarios,  noting that the relative energy densities of  the BLR, the dust fields, and the CMBR (at $z =0$) are, in units of erg cm$^{-3}$, 
$u_{BLR}\cong 0.026\phi_{-1}$, $u_{dust} \cong 3\times 10^{-3} (L_{IR}/10^{46}$ erg s$^{-1})/(R/{\rm pc)}^2$, and $u_{CMBR}\cong 4\times 10^{-13}$, respectively}.
We use the same parameters as derived by \cite{mal11} in our model.
 
\subsection{Target radiation fields}

Candidate target radiation fields are the cosmic microwave background radiation (CMBR), with dimensionless temperature $\Theta_{\rm CMBR}\equiv k_{\rm B}T_{\rm CMBR}/m_ec^2 \cong 4.6\times 10^{-10}(1+z)$, 1200 K and 660 K dust ($\Theta_{\rm dust1} \cong 2\times 10^{-7}$ and $\Theta_{\rm dust2} \cong 1.1\times 10^{-7}$, respectively), quasi-thermal scattered 10 eV disk emission ($\Theta_{\rm acr~disk} \cong 2\times 10^{-5}$). For Ly $\alpha$ line radiation---neglecting broadening, which is unimportant for these calculations---
$\e_{{\rm Ly}\alpha} = 10.2~{\rm eV} / m_e c^2 = 2\times 10^{-5}$. { Using the GZK energy of $\approx 6\times 10^{19}$ eV  as a yardstick where $\Delta$ isobar excitation by 2.725 K CMBR photons is a source of opacity and energy loss, the IR dust and accretion disk/Ly$\alpha$ radiation are effective targets for photopion production from cosmic-ray proton and neutron (nucleon) interactions with  $E_p \gg 10^{17}$ eV and $E_p \gg 10^{15}$ eV, respectively.} The energy densities of the target radiation fields are defined by the usual relation $u_0 \approx L/4\pi R^2 c$ where $R$ is the distance from the black hole.\footnote{This may underestimate the mean energy density inside a uniform spherical distribution of emitters, which is a factor 2.24 larger \citep{aa96}. We use, however, the simpler relation throughout.} 

{ The luminosity of photopion secondaries  is proportional to the effective photon number density for photopion production. For monochromatic line sources, $n_{eff} = u_0/ \epsilon_\star m_ec^2$, and for thermal blackbody or graybody radiation fields,  $n_{eff} = u_0/\Theta m_ec^2$, as shown below. Here, $\epsilon_{\star}$ is the mean energy of photons in the unit of $m_e c^2$. 
The effective photon number density 
for the CMBR is  
\begin{equation} 
n^{CMBR}_{eff}={u_{\rm CMBR} \over m_ec^2\Theta_{\rm CMBR}} \cong 
1.1 \times 10^{3} (1+z)^3 {\rm~cm}^{-3} \;
\label{uCMBRTheta}
\end{equation}
assuming the temperature of the CMBR at the present epoch is 2.725 K.
The effective photon density for $p\gamma$ scattering of the Ly $\alpha$ field is 
\begin{equation}
n^{Ly\alpha}_{eff}={u_{\rm Ly\alpha}\over m_ec^2 \epsilon_{\rm Ly\alpha } }  \cong 1.6\times 10^{9} \phi_{-1}{\rm~cm}^{-3} \;.
\label{uLyalphaTheta}
\end{equation}
The BLR radiation field due to the accretion-disk radiation scattered by 
BLR gas is, when approximated as a graybody, given by 
\begin{equation} 
n^{BLR}_{eff}={u_{\rm BLR}\over m_ec^2\Theta_{\rm acr~disk}}  \cong {1.7\times 10^9 \tau_{-1} \over  (\Theta_{\rm acr~disk}/2\times 10^{-5})  }{\rm~cm}^{-3} \;, 
\label{uacrdisk}
\end{equation}
assuming an effective Thomson scattering depth of $0.1\tau_{-1}$. For the two dust radiation fields, the effective photon densities for photopion p$\gamma$ interactions are 
\begin{equation}
n^{\rm dust1}_{eff}={u_{\rm dust1}\over m_ec^2\Theta_{\rm dust1}}  \cong {1.7\times 10^{10} L_{46} \over R_{\rm pc}^2  }{\rm ~cm}^{-3}
\label{udustTheta1}
\end{equation}
and
\begin{equation}
n^{\rm dust2}_{eff}={u_{\rm dust2}\over m_ec^2\Theta_{\rm dust2}}  \cong {3.1\times 10^{9} L_{45} \over  R_{\rm pc}^2}{\rm ~cm}^{-3}\;. 
\label{udustTheta2}
\end{equation}
Here the luminosity of the IR radiation, assumed to be radiated on a size scale of $R_{\rm pc}$ pc, is denoted $10^j L_j$ erg s$^{-1}$. 
To order of magnitude, $n_{eff} = u_0/m_ec^2\epsilon\sim u_0/m_ec^2\Theta \sim 10^9$ -- $10^{10}$ cm$^{-3}$, and the CMBR is negligible.

For sufficiently energetic cosmic-ray nucleons, the warm dust radiation field provides the densest and most important target photon field for photopion 
production. For dissipation of energy on the pc scale, only the highest energy nucleons with $E \gtrsim 10^{18}$ eV lose energy effectively by 
photohadronic processes with IR dust photons, because
the BLR radiation does not extend to the pc scale, and there is no other sufficiently 
strong radiation field at the pc scale to extract energy efficiently from these lower energy protons.}

\section{Model and Estimates} \label{sec:model}

The model is outlined, followed by a simple estimate of photohadronic efficiency and a simple derivation 
of secondary neutral production spectra made by photohadronic interactions of cosmic-ray jet protons in the inner jet.

\subsection{Model description}

This model is motivated by the hypothesis that that blazars and radio galaxies accelerate UHECRs \citep[e.g.,][]{mb92,bgg06,der09,mt09}; 
confirmation of this hypothesis is the detection of UHE neutrinos during blazar flares. Electromagnetic
signatures can also provide crucial evidence in support of UHECR production in blazars, and the presence of distinct hadronic emission 
signatures in the variable $\gamma$-ray SEDs is consistent with the SED shape but not unambiguous \citep[see \citet{2010arXiv1006.5048B} or, e.g.,][for Mrk 421]{2011ApJ...736..131A}. Detection of
a multi-TeV radiation signature in extreme high-synchrotron-peaked blazars like 1ES 0229+200 \citep{mdtm11} at energies where the EBL should suppress this emission  would support emission generated 
by UHECR protons accelerated by blazars \citep{ek10,ess10,ess11}.

\begin{figure}[t]
\begin{center}
 \includegraphics[width=3.0in]{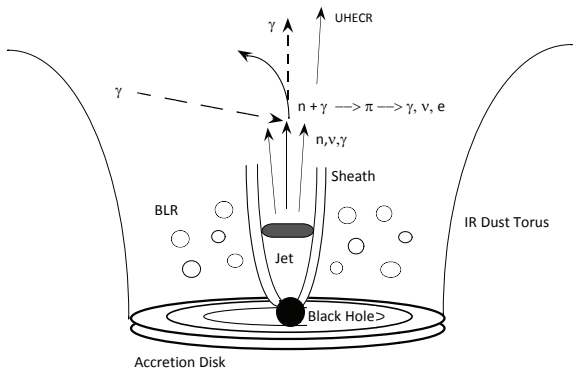} 
\vspace*{+0.2 cm}
 \caption{Cartoon of the system. Ultra-high energy neutrals, made through 
photohadronic production of cosmic-ray protons accelerated in the inner jet of 4C +21.35, escape to form a beam of neutrons,
neutrinos, and $\gamma$ rays. Subsequent photohadronic and $\gamma\gamma$ interaction of the neutrons and UHE $\gamma$ rays with IR photons from the dust torus induce a high-energy
lepton beam that quickly radiates secondary VHE photons to make the variable $\gamma$-ray emission in 4C +21.35. 
UHE neutrons and ions escaping from the system become UHECRs, and the jet radiation is accompanied by an associated flux of UHE neutrinos.
}
\label{cartoon}
\end{center}
\end{figure}

Fig.\ \ref{cartoon} shows a cartoon illustrating the underlying basis of this model. UHECR protons and ions are
accelerated in the outflowing relativistic plasma formed in the inner jet. Interactions with internal synchrotron 
and external quasi-isotropic radiation field cause ultra-relativistic protons accelerated in the outflowing
jet to undergo photopion losses, with the production of escaping neutrons, $\gamma$ rays, and neutrinos \citep{ad03}. Subsequent interactions of the escaping neutrals with IR torus photons on the pc scale make highly beamed leptons produced as secondaries of $\gamma\gamma$ interactions of UHE photons, and 
as secondaries of photopion interactions of UHE neutrons. The relativistic leptons make
beamed synchrotron and Compton scattered radiation. These emissions, being highly beamed, will vary on the production timescale
determined by the behavior in the inner jet.
Off-axis emission from misdirected leptons will not smear and enhance the timescale because it is produced by ultra-relativistic electrons that are not significantly deflected, so the contribution to the observer originates only from the particles traveling directly toward the observer.

\subsection{Radiation efficiency estimates}

We make simple estimates of the radiative efficiency for the production of 
photohadronic secondaries when UHE protons interact with photons from isotropic radiation fields external to the jet. In our estimates, the energy density of the external radiation field is denoted $u_{0}$
and the characteristic spatial size 
scale of the external radiation field is $R$. For thermal radiation fields, $\e_\star \cong 2.70\Theta$.

Suppose that a relativistic jet moving with bulk Lorentz factor 
$\Gamma$ is filled with cosmic-ray protons and ions. The comoving dynamical time for traveling through an external radiation field of  is $\tp_{\rm dyn}\cong R/\Gamma c$.
In the jet frame, the energy density of external radiation field is $u_0^\prime \cong \Gamma^2 u_0$, and 
the mean photon energy is $\ep_\star \cong \Gamma \e_\star$. The threshold Lorentz factor $\gp_{p,{\rm thr}}$ for photomeson production by protons with comoving Lorentz factor $\gpp$ interacting with photons with energy $\ep_\star$ 
is given by $\gp_{p,{\rm thr}} \approx \bar \e_{\rm thr}/\ep_\star$, where $\bar\e_{\rm thr} \approx m_\pi/m_e$ and 
$m_\pi \cong 140$ MeV is the pion mass.

{ The energy-loss rate of ultra-relativistic protons through photopion processes is therefore
\beq
t_{p\gamma}^{\prime -1} (\gpp>\gp_{p,{\rm thr}} ) \approx {(K_{p\gamma}\sigma_{p\gamma}) c u_0^\prime\over m_ec^2\ep_\star}\approx (K_{p\gamma}\sigma_{p\gamma}) c \Gamma({u_{0}\over m_ec^2  \e_\star}),
\eeq 
where $\hat\sigma \equiv K_{p\gamma}\sigma_{p\gamma}\cong 70~\mu$b  is the product of the inelasticity and the cross section for 
photomeson losses \citep{brs90,ad03}.  Note that the energy-loss timescale is proportional to the effective photon density $n_{eff} \cong u_0/m_ec^2\e_\star$.  Employing the approximation of \citet{ad03} here and in the following,} we take $\bar\e_{\rm thr} = 390$.  The photopion production efficiency is therefore simply
\beq
\eta_{p\gamma}(\gpp ) \cong \tp_{\rm dyn}/\tp_{p\gamma}(\gpp ) \approx \hat\sigma   R\,\left({u_{0}\over m_ec^2  \e_\star}\right)\;,
\label{etapgamma}
\eeq
provided $\gpp>\gp_{p,{\rm thr}}$. The Lorentz factor of protons measured in the stationary frame of the black-hole jet system (as if they had escaped the jet) is
$\gamma \cong \Gamma\gp$, so the threshold Lorentz factor is $\gamma_{p,{\rm thr}} \cong \bar\e_{\rm thr}/\e_\star$.

{ Following Equations (\ref{uLyalphaTheta}) -- (\ref{udustTheta2}), we write 
$n_{eff,q} = (u_0/m_ec^2\e_\star)_q \equiv [{(n_{eff}/ 10^q{\rm~cm}^{-3})}]$. 
For interactions in the BLR with Ly $\alpha$ photons or scattered accretion disk radiation, 
$\eta_{p\gamma} (\g_p )\approx 0.007 n_{eff,9} R_{17}$ 
for escaping protons with energy $E_p \gtrsim 2\times 10^{16}$ eV. 
Here, $R = 10^{17} R_{17}$ cm. For interactions with IR photons from the dust torus, 
$\eta_{p\gamma} (\g_p )\approx 0.07 n_{eff,10} R_{17}$ 
for protons with $E_p \gtrsim 7\times 10^{17}$ eV. 
For larger dissipation radii, one has $\eta_{p\gamma} (\g_p )\approx 2 n_{eff,10} R_{\rm pc}$, 
which means that UHECR protons in the jet plasma can dissipate essentially all of their energy 
into secondaries while traveling through the radiation field of the dust torus. But note that
in this case, the UHE proton beam has to be maintained throughout the entire pc-scale length of 
the external radiation field, in which case the emission would vary on timescales
no shorter  than $\sim R/\Gamma^2c \sim {10}^{4}~{\rm s} R_{pc} {(\Gamma/100)}^{-2}$.
}
%

 
Thus we see that the efficiency to produce photomeson secondaries through interactions of UHECR jet protons  with photons of an external radiation field is several \% in the inner jet, counting both line and scattered photons, and noting that $L_{\rm disk}\approx 5 \times 10^{45}$ erg s$^{-1}$ \citep{tan11}, so that $R_{\rm BLR}\approx 2\times 10^{17}$ cm, from Equation (\ref{RBLR}).
 This holds for $\gtrsim 10^{16}$ eV protons. Under optimistic conditions, 
the photopion efficiency of cosmic-ray protons in a blazar jet is of order unity in interactions with the IR dust photons, but this applies only to protons with escaping energies $\gtrsim 10^{18}$ eV.  Note furthermore that this is a conservative estimate of photohadronic efficiency,
 because it does not take into account internal synchrotron photons, or photons from 
a sheath-spine structure or from different portions of the jet \citep[compare][for TeV blazars, though similarly structured jets could be found in FSRQs]{gtc05,gk03} that could significantly enhance the 
target radiation field and the photohadronic efficiency.

It is interesting that all the $\Gamma$ factor dependencies in Equation (\ref{etapgamma})
drop out. This shows that we can equivalently calculate the photohadronic efficiency for a proton traversing the 
volume of the external radiation field without transforming to and from the comoving frame. We can furthermore make a simple derivation of the secondary neutron production by only considering interactions in the frame of the black hole and external radiation fields. 

{ With a photopion energy-loss cross section of $\approx 70~\mu$b, the column density of 
photons above threshold required for unity radiation efficiency is ${\cal N} \approx 1/(K_{p\gamma} \sigma_{p\gamma}) \sim 1.4\times 10^{28}$ cm$^{-2}$. The column density of Ly$\alpha$ photons in the BLR is
${\cal N}_{BLR}\sim n_{eff,Ly\alpha} R \sim 10^{26} n_{eff,9} R_{17}$ cm$^{-2}$. The column density of IR photons from the pc-scale warm dust torus is ${\cal N}_{dust1}\sim n_{eff,dust1} R \sim 3\times 10^{28} n_{eff,10} R_{pc}$ cm$^{-2}$. The  IR photon field  of the warm dust clearly presents the largest column density for photohadronic production, for the most energetic nucleons above the photopion reaction threshold. }

\subsection{Simple Derivation of Secondary Neutron Production Spectra}

The estimate above  implicitly assumes that when protons undergo photopion interactions, 
they remain protons. In about  50\% of the time, however, neutrons are formed which escape from the jetted plasma in a collimated outflow. The escaping neutrons deposit energy throughout the length of the jet by subsequent photohadronic interactions. We now estimate the energy loss rate of protons into neutrons via photopion production, and use that result to make a simple
derivation of the secondary neutron production spectrum, anticipating the more detailed derivation in the next section.

The inclusive energy-loss rate of protons into neutrons only is given by
\beq
-{dE_p\over dt}|_{p\gamma \rightarrow n} \;\cong\rho\chi\zeta\sigma_{p\gamma} E_p\;, 
\label{dEpdtpgamma}
\eeq
where $\rho$ is the cross-section ratio, $\chi$ is the fractional energy of the incident proton that is deposited into secondary 
neutrons, and $\zeta$ the neutron multiplicity. For neutron production near threshold, $\rho \cong 0.65$, $\chi \cong 0.8$, $\zeta\cong 1/2$, 
and the maximum photopion cross section is $\sigma_{p\gamma}= 520 \pm 30\, \mu$b. Here the proton energy $E_p = m_pc^2 \gamma_p$,
and $m_p \cong m_n$. 

The rate for a primary proton to lose energy into secondary neutrons is therefore simply
$$t^{-1}_{p\gamma\rightarrow n}(\gamma_p) = {|{dE_p\over dt}|_{p\gamma \rightarrow n}\over E_p} =$$
\beq
 {c\over 2}\int_0^\infty d\e\,n_{\rm ph}(\e ) \int_{-1}^1 d\mu (1-\mu)\rho(\bar\e_r) \chi(\bar\e_r) \zeta(\bar\e_r)\sigma_{p\gamma}(\bar\e_r )\;,
\eeq
where $\bar\e_r = \g_p\e(1-\mu)$ is the invariant interaction energy. Solving for a monochromatic radiation source $n_{\rm ph}(\e ) = u_0 \delta(\e-\e_\star)/ (m_ec^2\e)$
gives
$$-{dE_p\over dt}|_{p\gamma \rightarrow n} \;\cong$$
\beq 
{m_p\over m_e} c\Upsilon\sigma_{p\gamma} {u_0\over \e_\star}\,\gamma_p\,\left[ 1 - \left({\bar\e_{\rm thr}\over 2\g_p\e_\star}\right)^2\right]\,H\left(\g_p - {\bar\e_{\rm thr}\over 2\e_\star}\right)\;,
\label{dEpoverdt}
\eeq
where $\Upsilon \equiv \rho\chi\zeta$.\footnote{The Heaviside function with a single argument is defined by $H(x) = 1$ if $x \geq 0$ and $H(x) = 0$ otherwise. The second version of the Heaviside function is a step function defined by $H(x;a,b) = 1$ if $a < x < b$, and $H(x;a,b) = 0$ otherwise.}

The inclusive energy loss by primary protons with Lorentz factor $\g_p$ that is transformed into secondary neutrons with Lorentz factor $\gamma_n$ is conserved; therefore
\beq
-\int_1^\infty d\g_p \;{\left(d{\mathcal E}_p \over dt d\g_p \right)}_{p\gamma \rightarrow n} = \int_1^\infty d\g_n \;{d {\mathcal E}_n \over dt d\g_n},
\eeq
and $\g_p = \g_n/\chi$.
The neutron production luminosity is $L_{p\gamma \rightarrow n} = \g_p (d{\mathcal E}_p/dtd\g_p)_{p\gamma \rightarrow n}$.
For a stationary-frame distribution in the effective volume (where the retarded time is considered), $N_p(\g_p,\Omega_p)$ of protons differential in $\gamma_p$ and direction $\Omega_p$, the secondary production spectrum of neutrons, assuming that the neutrons have the same direction as the original protons, is 
$$\g_nL_n(\g_n,\Omega_n) = {m_p\over m_e}\; c\Upsilon \sigma_{p\gamma}{u_0\over \e_\star}\g_p^2 N_p(\g_p,\Omega_p) \,$$ 
\beq
\times\left[ 1 - \left({\bar\e_{\rm thr}\over 2\g_p\e_\star}\right)^2\right]\,H\left(\g_p - {\bar\e_{\rm thr}\over 2\g_p\e_\star}\right)\;.
\eeq

Using the method of \cite{gkm01,gkm04}, we transform the comoving jet frame proton spectrum $N^\prime_p(\g^\prime_p,\Omega^\prime_p)$ to the stationary frame jet proton spectrum $N_p(\g_p,\Omega_p)$ through the relation 
\begin{equation}
N_p(\g_p,\Omega_p) \equiv {d{\mathcal N}_p (\g_p,\Omega_p)\over d\g_p d\Omega_p} = \delta_{\rm D}^3\,N^\prime_p(\g^\prime_p,\Omega^\prime_p) = {\delta_{\rm D}^3\over 4\pi}\,N^\prime_p(\g^\prime_p)\;,
\label{NpgpOp}
\end{equation}
where the last relation is a consequence of the assumption of proton isotropy in the proper fluid frame, noting that $\g_p = \g_n/\chi = \dD \gpp$.
Thus
$$4\pi\g_n L_n(\g_n,\Omega_n) \approx {m_p\over m_e} {c\sigma_{p\g}u_0\over  \e_\star}\,\Upsilon \, \delta_{\rm D}^5\,[\g_{p}^{\prime 2} N^\prime_p({\g_{p}^\prime })]$$
\begin{equation}
 \times \, \left[ 1 - \left({\chi\bar\e_{\rm thr}\over 2\g_n\e_\star}\right)^2\right]\,H\left(\g_n - {\chi\bar\e_{\rm thr}\over 2\e_\star}\right)\;.
\label{gnlnapprox}
\end{equation}
Note the $\delta_{\rm D}^5$ beaming factor arising from the transformation of the proton spectrum, equation (\ref{NpgpOp}),  and one power each from energy and time.

From Equation (\ref{gnlnapprox}), we can determine the absolute amount of energy in protons, ${\cal E}_p = \Gamma{\cal E}_p^\prime$ that, while traveling through 
a radiation field characterized by $u_0/\e_{\star}m_ec^2$, will produce an apparent isotropic luminosity in neutrons $L_n = 10^{48}$ erg s$^{-1}$. 
If the spectrum of protons is $-2$, that is having equal energy per decade in protons, and extends well above the threshold $\g_{p,{\rm thr}} =\bar\e_{\rm thr}/ 2\e_\star $ for 
neutron production, then the fractional energy is proportional to the ratio of two logarithmic factors, which has a value of order $\approx 1/2$, implying
\beq
{1\over 2}\, {m_p\over m_e} {c\sigma_{p\g}u_0\over  \e_\star}\,\Upsilon\, {\delta_{\rm D}^5\over \Gamma}\,{{\cal E}_p \over m_pc^2} \gtrsim 10^{48} L_{48}\;{\rm erg~s}^{-1}\;.
\eeq

There is no fundamental reason why the engine should be Eddington-limited during outbursts, but this provides a fiducial amount of energy that could be generated by 
the central engine. Therefore we assume that the absolute nonthermal proton energy 
${\cal E}_p \approx 2 \times 10^{46} ~t_{\rm var} / (1+z) = 8.4\times 10^{48}$ erg, for the 4C +21.35 flare ($t_{\rm var} \approx 600$ s is the variability timescale of the flare and $z = 0.432$), which implies that 
\beq
{\delta_{\rm D}^5\over \Gamma}\gtrsim {1.2\times 10^{8}\over (\Upsilon/0.1) (u_0/\e_\star)_3}  
\eeq 
so $\dD \approx \Gamma \gtrsim 10^2$ in order to make an apparent luminosity $\gtrsim 10^{48}$ erg s$^{-1}$ in neutrons. 

The $\Gamma$ factor limit can be relaxed to $\dD \gtrsim 50$ if we happen to be viewing exactly down the jet ($\dD = 2\Gamma$) or if the energy content in protons is not limited by the Eddington luminosity during this outburst, but is instead one or two orders of magnitude larger. 
Alternately, small $\Gamma$ is possible if the photomeson production occurs mainly by target photons produced in inner jets (see below).  
In principle, there is no real reason that such large values of the Doppler factor or $\Gamma$ factor are not allowed. Arguments based on $\gamma\gamma$ opacity applied to the  July/August 2006 flares from PKS 2155-304 \citep{aha07} require $\dD\gtrsim 60$ \citep{bfr08},  while a full synchrotron/SSC model fit, including $\gamma\gamma$ and EBL effects, implies $\dD \gtrsim 100$ for PKS 2155-304 and $\dD \gtrsim 80$ for Mrk 501 with a variability timescale of $\approx 10^3$ s \citep{fdb08}. One-zone leptonic models for the VHE emission from 3C 279 \citep{alb08,ale11a} furthermore require large, $\Gamma \gtrsim 100$, bulk Lorentz factors \citep{brs09}.
Standard arguments relating variability time and location of emission site through the expression $t_{\rm var} \gtrsim (1+z) R/\Gamma^2 c$ imply $\Gamma \gtrsim 500$ for emission produced at the pc scale, which already undermines naive expectations about $\Gamma$ and the size of the emitting region. Measurements of superluminal motion never reach such values, but this is based on radio observations, which may be tracking the outflow speed of a radio-emitting sheath rather than a more rapidly moving spine. 
The necessity of a radical departure from standard models to explain the 4C +21.35 result seems unavoidable in all models \citep{tav11,nal12,tav12}; here we  construct a model within a framework that is testable by neutrino telescopes. 

\section{Spectral Powers and Beaming Factors for Photohadronic Secondaries } \label{sec:spec}

The system is now treated more carefully. 
A standard blazar jet scenario is considered \citep{ad01}. In the comoving frame defined by the existence of a quasi-isotropic particle and randomized magnetic-field distribution moving with Lorentz factor $\Gamma$ at some angle $\theta$ with respect to the observer, secondaries are formed as a result of photohadronic processes, e.g., neutrons (n), neutrinos ($\nu$), electrons and positrons  (e$^{\pm}$), photons ($\gamma$), protons (p), and $\beta$-decay electrons (e$_\beta$) and neutrinos ($\nu_\beta$). The charged secondaries may be trapped in the jet by its magnetic field, but the neutrals (n, $\nu$, and $\gamma$ rays) escape, forming a neutral beam with a beaming factor reflecting the Doppler factor of the jet and the production kinematics \citep{ad03}.

We start with Equation (2.44) of  \citet{dm09}, but instead of treating electron-photon scattering, here we treat secondaries of protons with Lorentz factor $\gamma_p$ made in photopion interactions with photons with dimensionless energy $\epsilon = h\nu/m_ec^2$. The technique is, however, general and can be adapted to other secondaries, noting that the dimensionless secondary energy $\epsilon_s$ in $m_ec^2$ units will be written as a fraction of $\chi$ of the primary proton energy $m_pc^2\gamma_p$. Thus for secondary photons, leptons or neutrinos, $\epsilon_s \cong \chi m_p\gamma_p/m_e$. For secondary neutrons, we can directly write, because $m_n \approx m_p$, the neutron production spectrum in terms
of the secondary neutron Lorentz factor $\gamma_n = \chi \gamma_p$. The direction-dependent emissivity for the production of neutrons through the process $p+\gamma \rightarrow n +X$ becomes 
 $$\g_n L_n(\g_n,\Omega_n) 
= m_pc^3\g_n^2\oint d\Omega
\int_0^\infty d\e\; {u(\e,\Omega)\over m_ec^2 \e}\;\oint d\Omega_p \;$$ 
\begin{equation}
\times (1-\cos\psi) \int_1^\infty 
d\gamma_p\;
N_p(\g_p,\Omega_p)\;
{d\sigma_{p\g}(\e,\Omega,\g_p,\Omega_p)\over d\g_n d\Omega_n}\;,
\label{gnln}
\end{equation}
specialized to ultra-relativistic protons, $\gamma_p\gg 1$. Here $\psi$ is the angle between the directions of the interacting proton and photon. The approximation we use for the differential cross section for $p\gamma$ photo-pion processes is a sum of step functions, given by 
$${d\sigma_{p\g}(\e,\Omega,\g_p,\Omega_p)\over d\g_n d\Omega_n} =$$
\begin{equation} 
\sigma_{p\gamma} \sum_{i=1}^N\zeta_i \rho_i 
H(\bar{\e}_r;\bar{\e}_i,\bar{\e}_{i+1}) 
\delta(\g_n -\chi_i\g_p) \delta(\Omega_n - \Omega_p )\;. 
\label{crosssection}
\end{equation}
Here the relativistic invariant is 
$\bar{\e}_r =\g_p\e(1-\cos\psi)$, 
$\cos\psi = \mu\mu_p+\sqrt{1-\mu^2}\sqrt{1-\mu_p^2} \cos(\phi-\phi_p )$, and 
$\arccos \mu $ and $\phi$ are the cosine angle and azimuthal angle, respectively, of the photon or (with subscript ``p") proton.  The term $\rho_i$ is the fractional cross section written as a ratio of the value of the step-function cross section to the maximum $p\gamma$ cross section $\sigma_{p\gamma} = 520 \pm 30 \,\mu$b. 
The effective multiplicity for segment $i$ between invariant energies $\bar\e_i \leq \bar\e_r < \bar\e_{i+1}$ is denoted $\zeta_i$, and $\chi_i$ is the mean energy fraction of a secondary neutron compared to the original proton energy for interactions associated with segment $i$.

Substituting Equation (\ref{crosssection}) into (\ref{gnln}) and solving gives 
$$\g_n L_n(\g_n,\Omega_n) 
= \left({m_p\over m_e}\right)\;c\sigma_{p\g}\g_n^2 \sum_{i=1}^N {\zeta_i\rho_i\over \chi_i}\oint d\Omega \;(1-\cos\bar \psi) $$ 
\begin{equation}
\times \int_0^\infty d\e\;{u(\e,\Omega)\over \e} N_p({\g_n\over \chi_i},\Omega_n)H(\bar{\e}^i_r;\bar{\e}_i,\bar{\e}_{i+1})\;, 
\label{gnln2}
\end{equation}
where 
$$\bar{\e}^i_r = {\g_n\over \chi_i}\e(1-\cos\bar\psi)\;$$ 
and $\cos\bar\psi = \mu\mu_n+\sqrt{1-\mu^2}\sqrt{1-\mu_n^2}\cos\phi$, taking $\phi_n = 0$ without loss of generality.

Using Equation (\ref{NpgpOp}), we have 
$$\g_n L_n(\g_n,\Omega_n)  
= \left({m_p\over m_e}\right)\;{c\sigma_{p\g}\over 4\pi }\, \delta_{\rm D}^3 \;\g_n^2   \sum_{i=1}^N {\zeta_i\rho_i\over \chi_i} $$ 
\begin{equation}
 \times \oint d\Omega \;(1-\cos\bar \psi) \int_0^\infty d\e {u(\e,\Omega)\over \e} \; N^\prime_p(\g^\prime_{pi})H(\bar{\e}^i_r;\bar{\e}_i,\bar{\e}_{i+1})\;,
\label{gnln3}
\end{equation}
where $\gp_{pi}\equiv {\g_n/ \chi_i\delta_{\rm D}}$.

We now specialize to the case of a surrounding external radiation field that is isotropic in the stationary frame of the black-hole/accretion-disk system. Therefore $u(\e,\Omega ) = u(\e )/4\pi$, and we can let $\mu_n = 1$ without loss of generality. After some straightforward manipulations, we obtain 
$$\g_n L_n(\g_n,\Omega_n) = {c\sigma_{p\g}m_p\over 4\pi m_e }\, \delta_{\rm D}^5 \;  \sum_{i=1}^N  \Upsilon_i \,[\g_{pi}^{\prime 2} N^\prime_p({\g_{pi}^\prime })] \times$$ 
\begin{equation}
\left[ \int_{y_i}^{y_{i+1}} d\e\; {u(\e)\over \e}- y_i^2\int_{y_i}^\infty d\e\; {u(\e)\over \e^3}+ y_{i+1}^2\int_{y_{i+1}}^\infty d\e\; {u(\e)\over \e^3}\right] \;,
\label{gnln4}
\end{equation}
where $\Upsilon_i \equiv \rho_i\zeta_i\chi_i$, and $$y_{i(+1)} \equiv {\chi_i\bar\e_{i(+1)}\over 2\gamma_n}\;.$$
Or, we have
$$\g_n L_n(\g_n,\Omega_n) = {c\sigma_{p\g}m_p\over m_e }\;  \sum_{i=1}^N  \Upsilon_i \,[\g_{pi}^{2} N_p({\g_{pi}, \Omega_n })] \times$$ 
\begin{equation}
\left[ \int_{y_i}^{y_{i+1}} d\e\; {u(\e)\over \e}- y_i^2\int_{y_i}^\infty d\e\; {u(\e)\over \e^3}+ y_{i+1}^2\int_{y_{i+1}}^\infty d\e\; {u(\e)\over \e^3}\right] \;.
\label{gnln4b}
\end{equation}

\subsection{External isotropic monochromatic radiation field}

For a monochromatic external radiation field, we write $u(\e) = u_0\delta(\e-\e_\star)$. In this case, 
$$\g_n L^{\rm line}_n(\g_n,\Omega_n) = \left({m_p\over m_e}\right)\;{c\sigma_{p\g}u_0\over 4\pi \e_\star}\, \delta_{\rm D}^5 \;  \sum_{i=1}^N  \Upsilon_i \,[\g_{pi}^{\prime 2} N^\prime_{p}({\g_{pi}^\prime})] \times$$ 
\begin{equation}
\left[ H(\e_\star;y_i,y_{i+1}) +{y_{i+1}^2\over \e_\star^2} H(\e_\star-y_{i+1}) -{y_i^2\over \e_\star^2} H(\e_\star-y_i) \right] \;.
\label{gnln5}
\end{equation}
This can also be written as 
$$4\pi\g_n L^{\rm line}_n(\g_n,\Omega_n) = {c\sigma_{p\g}u_0 m_p\over m_e\e_\star }\, \delta_{\rm D}^5 \;  \sum_{i=1}^N  \Upsilon_i \,[\g_{pi}^{\prime 2} N^\prime_p({\g_{pi}^\prime })] \times$$ 
$$\big[ H(\g_n;{\chi_i\bar\epsilon_i\over 2\epsilon_\star},{\chi_i\bar\e_{i+1}\over 2\e_\star}) +{y_{i+1}^2\over \e_\star^2} H(\g_n-{\chi_{i}\bar\epsilon_{i+1}\over 2\epsilon_\star})  -$$
\begin{equation}
  {y_i^2\over \e_\star^2} H(\g_n-{\chi_{i}\bar\epsilon_{i}\over 2\epsilon_\star}) \big ] \;.
\label{gnln6}
\end{equation}
The single step-function approximation with $\bar\e_2\rightarrow\infty $ takes the form 
$$4\pi\g_n L^{\rm line}_n(\g_n,\Omega_n) \approx {c\sigma_{p\g}u_0 m_p\over  m_e \e_\star}\,\Upsilon_1\, \delta_{\rm D}^5$$
\begin{equation}
 \times \,[\g_{p1}^{\prime 2} N^\prime_p({\g_{p1}^\prime })] \left[ 1 - ({y_1\over \e_\star})^2\right] H(\g_n-{\chi_1\bar\epsilon_{1}\over 2\epsilon_\star}) \;,
\label{gnln7}
\end{equation}
with $\Upsilon_1 = \chi_1\zeta_1\rho_1$, $\g_{p1}= \g_n/\chi_1\delta_{\rm D}$, and $y_1 = \chi_1\bar\epsilon_1/2\g_n$.
This confirms the approximate derivation, Equation (\ref{gnlnapprox}), and also shows that secondary production of photohadronic secondaries
has a beaming factor $\propto \delta_{\rm D}^5$.

\subsection{External graybody radiation field}

\begin{figure}[t]
\begin{center}
 \includegraphics[width=3.0in]{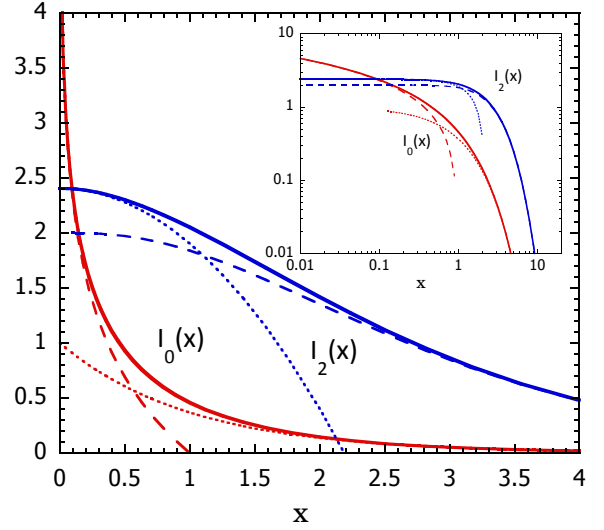} 
\caption{Functions $I_0(x)$ (red; Equation (\ref{I0w})) and $I_2(x)$ (blue; Equation (\ref{I2w})) with asymptotes (dashed and dotted lines).
Inset shows functions on a log scale.}
\label{I0(x)_I2(x)}
\end{center}
\end{figure}

The graybody radiation field, defined as a blackbody radiation field times the graybody factor $g$ giving the ratio of energy densities of the radiation field under consideration to that of a blackbody radiation field with the same effective temperature 
$T = m_ec^2\Theta/k_{\rm B}\;,$ 
is expressed as a spectral energy density in the form 
\begin{equation}
u_{\rm gb}(\e;\Theta) = g\,{8\pi m_ec^2\over \lambda_{\rm C}^3}\,{\e^3\over \exp(\e/\Theta )-1}\;.
\label{ugbeTheta}
\end{equation}
The electron Compton wavelength $\lambda_{\rm C} = h/m_ec = 2.42\times 10^{-10}$ cm. 
The energy density of blackbody radiation is 
\begin{equation}
u_{\rm bb}(\Theta ) = {8\pi^5 m_ec^2\over 15\lambda_{\rm C}^3} \,\Theta^4\;, 
\label{ubbTheta}
\end{equation}
and $g = u_0/u_{\rm bb}$, by definition.

Substituting Equation (\ref{ugbeTheta}) into (\ref{gnln4}) and solving gives 
$$4\pi\g_n L^{\rm gb}_n(\g_n,\Omega_n)  = {15 c\sigma_{p\g} m_p\over \pi^4 m_e}\;{u_0\over\Theta}\; \delta_{\rm D}^5 \;\sum_{i=1}^N  \Upsilon_i \,[\g_{pi}^{\prime 2} N^\prime_p({\g_{pi}^\prime })]  \times$$ 
\begin{equation}
 \left[I_2(x_i)-I_2(x_{i+1}) +x_{i+1}^2I_0(x_{i+1}) - x_i^2 I_0(x_i)\right] \;,
\label{gnln8}
\end{equation}
where 
\begin{equation}
x_{i(+1)} \equiv {\chi_i\bar\e_{i(+1)}\over 2\g_n\Theta}\;, 
\label{xi+1}
\end{equation}
$$I_0(x) \equiv \int_x^\infty {dw\over \exp(w)-1} =$$
\begin{equation}
 -\ln [1-\exp(-x)]\rightarrow 
\cases{- \ln x\;, &  $0 < x \ll 1$ \cr\cr 
\exp(-x)\;, &  $x \gg 1 $ \cr}\;, 
\label{I0w}
\end{equation}
and
$$I_2(x) \equiv \int_x^\infty {dw\,w^2\over \exp(w)-1} \rightarrow $$
\begin{equation}
\cases{2\zeta(3)-x^2/2\;, &  $x \ll 1$ \cr\cr 
(2+2x+x^2)\exp(-x)	 &  $x \gg 1 $ \cr}\;, 
\label{I2w}
\end{equation} 
where $\zeta(3) = 1.202\dots$ is Riemann's zeta function $\zeta(n)$ for $n=3$. Functions
$I_0(x)$ and $I_2(x)$ along with asymptotes are plotted in Figure \ref{I0(x)_I2(x)}.

In the one-segment approximation with $\e_2\rightarrow \infty$, 
$$4\pi\g_n L^{\rm gb}_n(\g_n,\Omega_n)  \approx $$
\begin{equation}
{15 c\sigma_{p\g} m_p \over m_e\pi^4 }\;{u_0\over\Theta}\; \delta_{\rm D}^5 \; \Upsilon_1 \,[\g_{p1}^{\prime 2} N^\prime_p({\g_{p1}^\prime })]  I(x)
\;, 
\label{gnln9}
\end{equation}
where $x = \chi_1\bar\epsilon_1/ 2\g_n\Theta$, and 
\begin{equation} 
I(x) \equiv I_2(x)- x^2 I_0(x)  \rightarrow 
\cases{2\zeta(3) + x^2(\ln x - 1/2)\;, &  $x \ll 1$ \cr\cr
(2 + 2x)\exp(-x) &  $x \gg 1 $ \cr}\;. 
\label{I2x2I0x}
\end{equation} 
Integrals $I_i(x) = \int_x^\infty dw w^i/[\exp(w)-1]$ also arise in the 
study of anisotropic Compton scattering of stellar blackbody radiation \citep{db06}. 
The large-$x$ asymptote  is accurate to within 20\% even for small $x$ (Fig.\ \ref{I0(x)_I2(x)}).

\subsection{Secondary photopion production cross section}

An $N = 2$ segmented cross section was used in \cite{ad03} and \cite{dm09}, which accounted for $\Delta(1232)$ resonance and multi-particle production in $p+\gamma\rightarrow \pi+X$ interactions. We extend the $N=2$ segment approximation to $N = 3$, which now better accounts for the production of heavy resonances. Parameters are given in Table \ref{table:parameters}. The number of segments in the approximation can be made arbitrarily large, but energy dispersion, amounting to another integration, is required to reproduce the accuracy of Monte Carlo simulations. The minor loss in accuracy is compensated by code flexibility. 
Detailed numerical calculations of neutrinos in relativistic jet sources, where the multi-pion production is taken into account, are found in e.g., \cite{2001ICRC....3.1153M,mur07}. 

\begin{table}[t]
\begin{center}
\caption{$N=3$ segment approximation parameters for neutrons and neutrinos formed as secondaries in photopion interactions and from neutron beta decay. Parameters are invariant energies $\bar\e_i$, cross-section ratio $\rho$, fractional energy $\chi$, and multiplicity $\zeta$.}
\begin{tabular}{cclllllll}
\hline
 &   &   & n  &  & $\nu$ &    & $\nu_\beta$ &  \\
$i$ & $\bar\e_i$ & $\rho_i$ & $\chi_i^n$ & $\zeta^n_i$ &  $\chi^\nu_i$ & $\zeta^\nu_i$ &  $\chi^\nu_i$ & $\zeta^\nu_i$ \\ 
\hline
\hline
 &  &  &   &   &  & &  &   \\ 
1 & 390 & 0.65 & 0.8 & 1/2 & 0.05 & 3/2 & 0.001 & 0.5\\
 &  &  &  & & &   & &   \\ 
2 & 980 & 0.44 & 0.6 & 1/2 & 0.05 & 3  & 0.001 & 0.5 \\
 &  &  &   &  & &   & &  \\ 
3$^a$ & 2000 & 0.23 & 0.4 & 1/2 & 0.05 & 6  & 0.001 & 0.5 \\
\hline\end{tabular}
\label{table:parameters}
\end{center}$^a$ In statistical multi-pion production at high energies, 
smaller values, e.g., $\chi_3^\nu\approx 0.03$ can be found via detailed simulations \citep[e.g.,][]{muc+00,mur07}, 
with correspondingly higher multiplicity, such that the product  $\chi_3^\nu\zeta_3^\nu $ remains at about the same value as given here. 
\vskip0.2in
\end{table}

\subsection{Jet power and energetics}

The apparent isotropic jet power during a short flaring episode of a black-hole jet can be large, and must be $> 10^{48}$ erg s$^{-1}$ in order
to reproduce the value of $L_\gamma \approx 10^{47}$ erg s$^{-1}$ observed in VHE $\gamma$-rays from 4C +21.35 during the flare (the 
GeV $\gamma$-rays can, and are likely to be, produced in the inner jet; see Section 7). For a continuous jet, which makes 
persistent emission over long periods of time, 
the jet power is likely to be smaller, but still has to exceed $\approx 2.5\times 10^{47}$ erg s$^{-1}$, which is the two-year time-averaged
$\gamma$-ray luminosity measured with the {\it Fermi}-LAT for 4C +21.35 \citep{2011ApJ...743..171A}.

For a continuous jet,  the total jet power supplied by the black hole  is composed of three terms from the magnetic-field, particle, and photon power. In the naive 
standard model, we write the total jet power for a two-sided jet as 
$${\cal P}_\star = {\cal P}_{\star,B} +{\cal P}_{\star,{\rm par}}+ {\cal P}_{\star,\g}  =$$
\begin{equation}
 2\Omega_j\beta c R^2 \Gamma^2 \left[ \left( {B^{\prime 2}\over 8\pi }\right)+ {{\cal E}^\prime_{\rm par}\over V^\prime_b}\right] + {8\Gamma^2\over 3\delta_{\rm D}^4}L_{\rm syn} + {32\Gamma^4\over 5\delta_{\rm D}^6}L_{{\rm EC}}
\label{Pstar}
\end{equation}
 \citep{1993MNRAS.264..228C,2008MNRAS.385..283C}. 
The final two terms in this expression represent the photon power from synchrotron and Compton processes, respectively, and are derived in Appendix \ref{app:A}.
Here $L_{\rm syn}$ and $L_{\rm EC}$ are the measured apparent bolometric luminosities of the synchrotron and Compton processes, respectively, noting
that the synchrotron term applies to emission that is isotropic in the comoving frame with beaming factor $\delta_{\rm D}^4$, and so 
would also apply to synchrotron-self Compton emission, while the Compton term applies to the external Compton process with 
beaming factor $\delta_{\rm D}^6$, which applies both in the Thomson \citep{der95} and Klein-Nishina regimes \citep{gkm01,gkm04}.

The comoving spherical blob volume is, of course, $V^\prime_b = 4\pi r_b^{\prime 3}/3$, 
and the particle power, assumed to be dominated by the hadrons (in particular, the accelerated protons), is just
\begin{equation}
{\cal E }_{\rm par}^\prime = m_pc^2 \int_1^\infty d\gp \,\gp\,N_p^\prime (\gp )\;.
\label{Uparprime}
\end{equation}
The jet opening solid-angle $\Omega_j \cong \pi r_b^{\prime 2}/ R^2$. 

The comoving magnetic field $B^\prime$ required to accelerate particles to energy $E$ after escaping the jet is
restricted by the \citet{hil84} condition, namely that the comoving particle Larmor radius $r^\prime_{\rm L} = E^\prime/QB^\prime \cong E/\Gamma Z e B^\prime < r_b^\prime$.
 The causality condition restricts the blob radius to be
\begin{equation}
r_b^\prime \lesssim {c\delta_{\rm D} t_{\rm var}\over 1+z}\;,
\label{rbprime}
\end{equation}
where the  measured variability time is denoted by $t_{\rm var} \equiv 10^3 t_{\rm var,3}$ s. Thus the minimum magnetic field $B^\prime_{\min }$ required for proton acceleration to energy $10^{20}E_{20}$ eV is given by
\begin{equation}
B^\prime_{\min} = {(1+z) E\over \Gamma e c \delta_{\rm D} t_{\rm var}}\cong   \;{E_{20} (1+z)\over (\Gamma/100)(\delta_{\rm D}/100)\, t_{\rm var,3}}\;{\rm G}. 
\label{Bprimemin}
\end{equation}

The accelerated proton distribution in the fluid frame is assumed to take the form
\begin{equation}
\g_p^{\prime 2} N^\prime_p({\g_p^\prime }) = K^\prime H(\g_p^\prime - \g^\prime_{\min}) \g_p^{\prime 2-s}\exp(-{\g_p^\prime\over \g_{\max}^\prime } )\;,
\label{gpp2npp}
\end{equation}
so that $\gamma^\prime_{\max} \cong   {1.1\times 10^{11} E_{20} /\Gamma}$ 
when $B^\prime = B^\prime_{\rm min}$.
The normalization for $K^\prime$ in terms of the jet-frame particle energy content 
\begin{equation}
{\cal E}^\prime_{\rm par} \cong {m_pc^2 K^\prime \gamma_{\rm min}^{\prime 2-s}\over s-2 }\;[ 1-({\gp_{\rm max}\over\gamma_{\rm min}^{\prime}})^{2-s}]\; \stackrel{s\rightarrow 2}{\rightarrow }\; m_pc^2 K^\prime \ln G 
\;,
\label{Uparprime1}
\end{equation}
where $G\equiv \g_{\rm max}^\prime/\g_{\rm min}^\prime $. 
The total particle power, assumed to be dominated by accelerated protons, is from Equation (\ref{Pstar}) given by 
\begin{equation}
{\cal P}_{\star,{\rm par}} = {3\over 2 r_b^\prime } \,\beta c \Gamma^2 {\cal E}^\prime_{\rm par}\;.
\label{calPstarpar}
\end{equation}
The total magnetic-field power ${\cal P}_{\star,B} = \beta c r_b^{\prime 2} \Gamma^2 ( {B^{\prime 2}/4 } )$. Using Equation (\ref{Bprimemin}) with $B^\prime \cong B^\prime_{\rm min}$ gives
\begin{equation}
{\cal P}_{\star,B} = {c\over 4}{E^2\over Z^2 e^2} \cong 8.3\times 10^{44}\beta  {E_{20}^2\over Z^2}\;{\rm erg~s}^{-1}\; 
\label{calPstarB}
\end{equation}
for relativistic outflows, which recovers a familiar result \citep{wax04,fg09,dr10}. Note that unlike the magnetic-field power, 
the particle power ${\cal P}_{\star,{\rm par}}\propto r_b^{\prime -1}$, and becomes larger for smaller $t_{\rm var}$.

\subsection{Maximum Particle Energy and Bulk Lorentz Factor}

The minimum bulk Lorentz factor from $\gamma\gamma$ attenuation is given by $\Gamma_{\rm min} \cong [\sigma_{\rm T} d_L^2 (1+z)^2 f_{\hat \e} \e_1/4t_{\rm var} m_ec^4]^{1/6}$, where $f_\e = 10^{-12}f_{-12}$ erg cm$^{-2}$ s$^{-1}$ is the $\nu F_\nu$ flux of the target photons (assumed to be produced co-spatially with the highest energy photon with energy $m_ec^2\e_1$) evaluated at  $\hat \e = 2\Gamma_{\rm min}^2/(1+z)^2\e_1$. For 4C +21.35, 
\begin{equation}
\Gamma_{\rm min} \cong 14.1\; \left[ {f_{-12} (E/100{\rm~GeV})\over (t_{\rm var}/600{\rm~s}) }\right]^{1/6}\;.
\label{Gammamin}
\end{equation}
Photons measured with energies of 100 GeV preferentially interact through $\gamma\gamma$ processes with photons with energies $\hat E = m_ec^2 \hat \e \cong 0.5$ keV, where $f_{-12} \cong 1$ \citep{tav11}. 
This estimate assume co-spatiality of the emission, and only takes into account $\gamma\gamma$ absorption by photons in the jet. 

The \citet{hil84} condition, with $B^\prime$ given by Equation (\ref{Bprime}) and the emission region size scale $r_b^\prime$ given by Equation (\ref{rbprime}), implies a maximum energy 
$$E_{\rm max} \cong Ze\Gamma B^\prime r_b^\prime \cong $$
\begin{equation}
4 \times 10^{20} Z \left({\Gamma\over 100} \right)^3 \left( \frac{t_{\rm var}}{(1+z)600{\rm~s}}\right)\;\sqrt{\phi_{-1}\over (A_{\rm C}/100)}\;{\rm~eV}\;
\label{EmaxH}
\end{equation}
of particles accelerated by and escaping from the jet. 
Here $\dD \approx \Gamma$ is used. 
Thus, 
though other losses such as photohadronic cooling may be relevant, 
acceleration of UHECR protons is feasible in the inner jet of 4C +21.35 provided $\Gamma \gtrsim 100$ 
and $B^\prime \approx 10$ G. 
There is no conflict between this value and Equation (\ref{Gammamin}).

\subsection{Single blob versus continuous jet}

In our formulation of the problem, the secondary production spectrum was derived for
a single {\it one-zone blob}, but the jet power was
derived for a {\it continuous jet}.  The energy density in the blob, 
which is mainly in the form of energetic particles, is equated with 
the product of the energy density used to determine the power of the continuous jet and
the blob volume determined through Equation (\ref{rbprime}).
Consequently the single blob represents only a single slice with comoving width $r^\prime_b$ and stationary
frame width $\approx c t_{\rm var}/(1+z)$ of the continuous jet, and has a comoving 
particle energy content given by Equation (\ref{calPstarpar}). 

This is not, however, the only or the most intuitive
normalization. As outlined in the estimates, Sections 3.2 and 3.3, we can alternately constrain the total energy in an outburst that lasts
for $t_{\rm var}$ to be Eddington-limited, so that the total particle energy ${\cal E}_{\rm par} \lesssim L_{\rm Edd} t_{\rm var}/(1+z)$, 
or the comoving particle energy 
\beq
{\cal E}^\prime_{par} \lesssim L_{\rm Edd} t_{var}/\Gamma(1+z).
\label{Eprimepar}
\eeq
Comparing
with Equation (\ref{calPstarpar}) and requiring ${\cal P}_{\star,{\rm par}} \leq L_{\rm Edd}$ shows that the two normalizations
differ by a factor $2\dD/3\Gamma$, which is of order unity for emission within the Doppler beaming cone.

We first derive the neutron production luminosity for a single blob traveling through the external radiation field. Because it represents only one zone of the continuous jet, it can severely underestimate the possible secondary power for a persistent jet.
Afterword we consider secondary emission from a continuous jet, noting that this produces a nonvariable/continuous outflow with duration of $\approx (1+z) R/c$, whereas the single blob gives an outgoing pulse of secondaries that preserves the engine variability timescale.

\subsubsection{Discrete blob}

The apparent isotropic luminosity of secondary neutron produced by cosmic-ray protons in a jet traveling through a background
external isotropic monochromatic radiation field characterized by $u_0/\e_\star$ is,  from Equation (\ref{gnln7}),
given  by the expression 
$$4\pi L_n^{\rm line}(\Omega_n) = 
4 \pi \int_1^\infty d\g_n \;L^{\rm line}_n(\g_n,\Omega_n) \cong$$
$$ {c\sigma_{p\g} }\,{m_pu_0\over m_e\epsilon_\star}\,{\Upsilon_1 \, \delta_{\rm D}^5 K^\prime\over s-2} \; \left( {\bar\e_1\over 2\dD\e_\star }\right)^{2-s} 
\left[ 1 - \left( {2\dD\e_\star\g_{\rm max}^\prime\over \bar\e_1} \right)^{2-s}\right]$$ 
\begin{equation}
 \stackrel{s\rightarrow 2}{\rightarrow }\; {c\sigma_{p\g} }\,{m_pu_0\over m_e\epsilon_\star}\,\Upsilon_1\, \delta_{\rm D}^5 K^\prime \; \ln\left( { 2\dD\e_\star\g_{\rm max}^\prime\over \bar\e_1 }\right)
\;,
\label{gnln11}
\end{equation}
restricting to the $s\rightarrow 2$ limit in the final expression. 
Here, the term $[1 -  (y_1 / \epsilon_\star)^2]$ in Eq. (\ref{gnln7}) is neglected because it makes only a small correction to the integral.

If a single blob explains the 4C +21.35 observations, 
then  a minimal requirement is that the apparent secondary power in neutrons exceeds the radiant power $L_\gamma = 10^{48}L_{48}$ erg s$^{-1}$ measured in the observer direction, and which is subsequently converted to an apparent isotropic luminosity of $\approx 10^{47}$ erg s$^{-1}$ in $\g$ rays at the pc scale. The condition $4\pi L_n^{\rm line}(\Omega_n) \geq 10^{48} L_{48}$ erg s$^{-1}$ implies, when $s = 2$,  
\begin{equation}
K^\prime_{\rm line}\gtrsim {3.6\times 10^{59}L_{48}\over \Upsilon_{1,-1} (u_0/\e_\star)_3\dD^5 \ln A}\;,
\label{Kdef}
\end{equation}
where 
$\Upsilon_1 \equiv 10^{-1} \Upsilon_{1,-1}$, 
and $ A\equiv \left( { 2\dD\e_\star\g_{\rm max}^\prime/\bar \e_1 }\right)$. 
From Equations (\ref{Uparprime1}), (\ref{calPstarpar}) and (\ref{Kdef}), the absolute jet power is required to be at least 
\begin{equation}
{\cal P}^{\rm line}_{\star,{\rm par}} > { 8\times 10^{45}\ln G\over t_{\rm var,3}\ln A}\;{(\Gamma/100)^2\over (\dD/100)^6}\; {(1+z)L_{48}\over (u_0/\e_\star)_3\Upsilon_{1,-1}}\;{\rm erg~s}^{-1}\;.
\label{calPstarpar1}
\end{equation}

The analogous expressions for a thermal graybody radiation field, from Equation (\ref{gnln8}), is 
$$4\pi L_n^{\rm gb}(\Omega_n) \cong {30\over \pi^4}\,{m_pc\sigma_{p\g} u_0\over m_e\Theta}\,{K^\prime\Upsilon_1\, \delta_{\rm D}^{3+s} \over s-2} \left( {\e_1\over 2\Theta }\right)^{2-s}\times$$
\begin{equation}
 \left[ 1 - \left( {2\Theta\dD\g_{\rm max}^\prime\over \bar\e_1} \right)^{2-s}\right]\stackrel{s\rightarrow 2}{\rightarrow }\; {15 c\sigma_{p\g}\over 2\pi^5 }\,{m_pu_0\over m_e\Theta}\,K^\prime \Upsilon_1\, \delta_{\rm D}^5\;\ln C\;
\;,
\label{gnln12} 
\end{equation}
where 
$C \equiv \left( { 2\Theta \dD\g_{\rm max}^\prime/ \bar\e_1 }\right)$, implying 
\begin{equation}
K^\prime_{\rm gb}\gtrsim {1.2\times 10^{59}L_{48}\over \Upsilon_{1,-1}[u_0/\Theta]_{4} \dD^5 \ln C}\;.
\label{Kdef2}
\end{equation}
From Equations (\ref{Uparprime1}) and (\ref{calPstarpar}), the absolute jet power when $s = 2$ is at least
\begin{equation}
 {\cal P}_{\star,{\rm par}} > {3\times 10^{45}(1+z) L_{48}\;(\Gamma/100)^2 \ln G\over t_{\rm var,3}(u_0/\Theta)_{4}\Upsilon_{1,-1}(\dD/100)^6\ln C} \;{\rm erg~s}^{-1}\;.
\label{calPstarpar2}
\end{equation}
Using the alternate normalization for the single blob comoving energy content, Equation (\ref{Eprimepar}), amounts to multiplying these estimates by the  $2\dD/3\Gamma$ factor with no impact on these jet-power estimates 
for jets observed within the Doppler beaming cone.

Equations (\ref{calPstarpar1}) and (\ref{calPstarpar2}) can be read as saying that to produce apparent isotropic neutron luminosity of $\approx 10^{48}$ erg s$^{-1}$ from an outburst lasting $\sim 10^3$ s to the observer requires 
a jet with $\dD\approx \Gamma \approx 10^2$ and absolute jet power of  $\sim 10^{46}$ erg s$^{-1}$ primarily in the form of cosmic-ray protons with an $s=2$ spectrum, 
confirming the estimates of Section 3.  The large Doppler factors 
and associated jet collimation, with jet opening angle $\theta\sim 1/\Gamma$, overcome the photohadronic inefficiency and shortens the observed timescale of emission to
be consistent with sub-Eddington luminosities and rapid variability.

\subsubsection{Continuous jet}

The differential secondary neutron production luminosity $\g_n L_n(\g_n,\Omega_n)$  applies to a single blob traveling through a background 
external radiation field of spatial extent $ R$. This emission persists for a time $\Delta t =  R(1+z)/\beta\Gamma\dD c$ for an 
observer at angle $\theta$ with respect to the jet axis, so we can write the time-dependent secondary production spectrum for a 
single zone as
$\g_n L_n(\g_n,\Omega_n,t) = \g_n L_n(\g_n,\Omega_n) H(t; t_0, t_0 + \Delta t)$, where $t_0$ is some arbitrarily chosen zero of time.

For a succession of blobs that comprise the continuous jet, the time-averaged spectral luminosity is 
$$\langle \g_n L_n(\g_n,\Omega_n)\rangle = (\rho_*\Delta t)\; \g_n L_n(\g_n,\Omega_n)\lesssim$$
\beq
  {(1+z)^2  R\over c t_{\rm var} \Gamma\dD } \;\g_n L_n(\g_n,\Omega_n)\;,
\label{langlegnLnrangle}
\eeq
  where $\rho_*$ is the rate at which blobs are ejected. The largest possible value of $\rho_*$  corresponds to the ejection 
of blobs at the rate $\beta c/\Delta r$, where  $\Delta r = c t_{\rm var}/(1+z)$ is the stationary-frame width of the blob, that is, when the time between blob ejection is just $\Delta r/\beta c$, corresponding to the continuous jet limit. This explains the coefficient in the final term of Equation (\ref{langlegnLnrangle}). { There is reduction by one power in the beaming factor from a 
discrete blob to a continuous jet, as is well-known \citep{lb85}.}

\subsubsection{Internal synchrotron photons}

An alternative to assuming large Doppler factors is to have small Doppler factors 
($\dD \ll 100$), so that the internal synchrotron photons present a dense target radiation field \citep{ad01,ad03}; see Appendix \ref{app:B}.  But the Doppler amplification of the received luminosity would be much reduced by such small Doppler factors, so that absolute jet powers larger than the Eddington luminosity
would be required.  Consequently, the large Doppler factor solution seems preferable unless an alternative mechanism to make rapid variability at the pc scale is devised. 
Note furthermore that photons from the direct accretion-disk radiation field \citep{2001ICRC....3.1153M}, which provide an additional important target photon source that has not been considered here, is more important for jetted emitting regions far from the black hole when the Doppler factor is larger, because aberration of accretion-disk photons is greater
\citep{ds02}.

\subsection{Calculations}

We use the formalism developed here to make calculations of secondary neutron and neutrino production spectra from the interactions between a power-law  distribution of UHECR protons, isotropic in the jet rest frame,  that interact with BLR radiation in the inner jet to make secondary neutrons, pions, and neutrinos. The results are expressed in the form 
$4\pi E L(E,\Omega) = 4\pi E d^2 L(E,\Omega)/ dE d\Omega $ 
(units of erg s$^{-1}$), so as an apparent isotropic luminosity for the production of secondaries with that energy. Given that we are scaling to 4C +21.35 with $4\pi d_L^2 = 6.52\times 10^{56}$ cm$^2$ and bolometric energy fluxes during flares exceeding $10^{-9}$ erg cm$^{-2}$ s$^{-1}$, this corresponds to apparent bolometric $\gamma$-ray luminosities $\gtrsim 10^{48}$ erg s$^{-1}$.

\begin{figure}[t]
\begin{center}
 \includegraphics[width=3.0in]{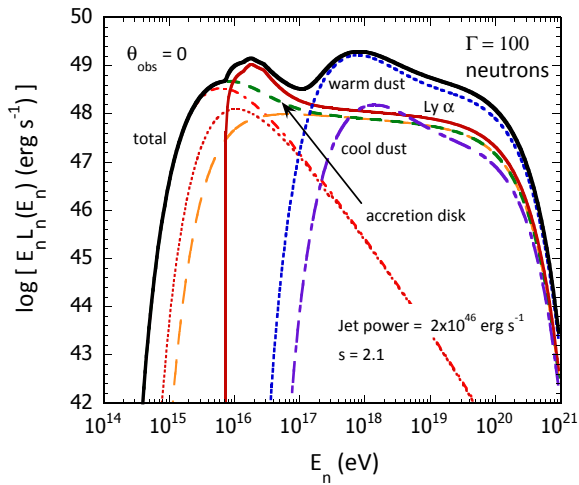}
 \caption{{ Black solid curve shows the secondary production spectrum of {\it neutrons} made from cosmic-ray protons accelerated in the inner jet of a blazar. The jet is assumed to have bulk Lorentz factor $\Gamma = 100$, be aligned along the jet axis ($\theta_{\rm obs} = 0^\circ$, $\dD = 200$), and interact with BLR and torus photons. The separate components  from interactions with Ly$\alpha$, scattered accretion disk, and warm and cool dust radiations are labeled. The contributions from the different resonance and multipion interactions are shown separately for the scattered accretion-disk radiation component. } } 
\label{nLnangle}
\end{center}
\end{figure}

\begin{figure}[t]
\begin{center}
 \includegraphics[width=3.0in]{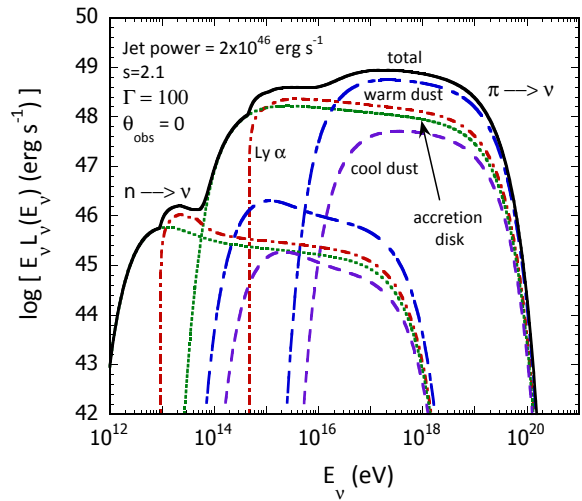} 
 \caption{{ Black solid curve shows the secondary production spectrum of {\it neutrinos} made from cosmic-ray protons accelerated in the inner jet of a blazar, using the same parameters as in Fig.\ \ref{nLnangle}. The separate components for neutrino production from interactions with Ly$\alpha$, scattered accretion disk, and warm and cool dust radiations are labeled for the channel $(p+\gamma \rightarrow) \pi \rightarrow \nu$. The corresponding line styles are shown for the production
of $\beta$-decay neutrinos denoted by $n\rightarrow\nu$, which is formed when the neutrons, assumed to escape the production site and travel rectilinearly until decay, make neutrinos through neutron $\beta$ decay $n \rightarrow p +e^- +\overline\nu_e$.   } 
}
\label{nLnGamma}
\end{center}
\end{figure}

{ Flaring results to model the SED of 4C +21.35 from photohadronic neutron and neutrino emissions made by UHECR particle acceleration in the inner jet are shown in Figures \ref{nLnangle} and \ref{nLnGamma}, respectively. In the picture studied here, cosmic-ray protons in a single-zone blob with $\Gamma = 100$ and $\dD = 200$---so looking directly down the jet---interact  with photons of the BLR and the dust torus. } Here we normalize to a total jet power ${\cal P}_{\star}=  2\times 10^{46}$ erg s$^{-1}$, which is the Eddington luminosity of a $1.5\times 10^8 M_\odot$ black hole. We use $t_{\rm var} = 600$ s for the variability time scale, and  Equation (\ref{Eprimepar}) to normalize the energy content of jet protons.  This jet energy is fed into magnetic field, photons, and particles, limited by ${\cal P}_{\star} = L_{\rm Edd}$. { To accelerate protons to $10^{20}$ eV,  $B^\prime \gtrsim 1$ G, from Equation (\ref{Bprimemin}).}
Most of the power goes into particles; the magnetic field required to accelerate $10^{20}$ eV protons leads to a total jet magnetic-field power of  $\approx 10^{45} E_{20}^2$ erg s$^{-1}$, from Equation (\ref{calPstarB}), which is $\approx 5$\% of $L_{\rm Edd}$. The photon power is smaller still, due to the small beaming factor $\approx 1/\Gamma^2$. We let $\gp_{\rm min} = \Gamma$, corresponding to the minimum Lorentz factor of protons swept into the jet. The effective photon densities for the Ly $\alpha$ and scattered accretion disk radiation field are given by Equations (\ref{uLyalphaTheta}) and (\ref{uacrdisk}), respectively. The total power that is not in magnetic field or photons is assumed to be mostly in the form of a nonthermal particle distribution of the form of Equation (\ref{gpp2npp}), with $s = 2.1$.

{ Figure \ref{nLnangle} shows the neutron production spectrum for a head-on $\Gamma = 100$ jet, $\theta_{\rm obs} = 0^\circ$. The large apparent neutron luminosities, exceeding $10^{49}$ erg s$^{-1}$ from an AGN with absolute jet power of just $2\times 10^{46}$ ergs, shows that photohadronic processes in the inner jet can be very efficient \citep[contrary to][]{nal12}, with  the warm dust  more effective than  accretion-disk radiation for hard cosmic-ray proton spectrum that extends to $10^{20}$ eV.  The two peaks found in the production/injection of neutrons into the pc-scale region where only IR dust photons are present means that only the highest energy neutrons with $E_n\gtrsim 10^{18}$ eV, are effective at making photopion secondaries beyond the inner jet within the BLR.}

{ Figure \ref{nLnGamma} is a calculation of the  neutrino production spectra for an inner-jet system with the
same parameters as in Fig.\ \ref{nLnangle}.  No mixing corrections are made. The apparent bolometric luminosity is $\gg 10^{48}$ erg s$^{-1}$, which as noted before, requires large Lorentz factors and preferential Doppler factors, and production in the inner jet for  short variability. The  components labeled by $\pi \rightarrow \nu$ refer to neutrino production in photopion interactions, $p+\gamma \rightarrow \pi \rightarrow \nu$.  The low-energy feature in the neutrino production spectra labeled $n\rightarrow\nu$ are $\beta$-decay electron neutrinos, when the bulk of the neutrons escape and decay in intergalactic space. Both the neutron and neutrino spectra become very steep, harder than $+2$ in $E L(E)$, below PeV energies, as a result of threshold effects.   Detailed calculations \citep[e.g.,][]{mur07} find softer spectra below PeV energies than calculated here due to the spread in the energy distribution of low-energy pions, especially from $\pi$/$\mu$ decay kinematics and partially from multi-pion production. }

\begin{figure}[t]
\begin{center}
 \includegraphics[width=3.0in]{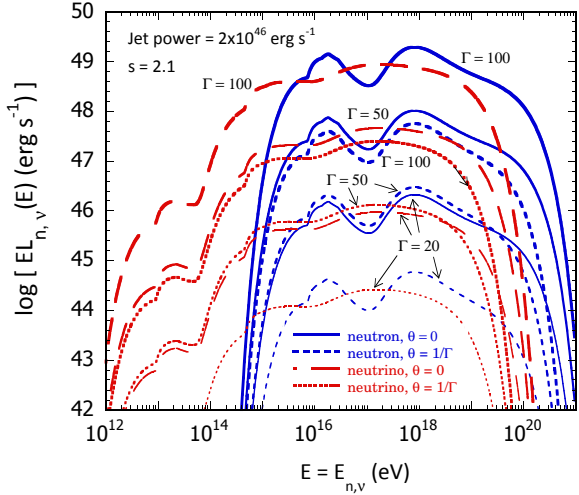} 
 \caption{{ Comparison of production spectra of neutrons (blue curves) and neutrinos (red curves) for jets with different Lorentz factors and observing angles, as labeled. For the standard inner jet and accelerated cosmic-ray spectrum described in the text, the production spectra are calculated for $\Gamma = 100, 50$ and 20 at 
observation angles $\theta = 0$ and $\theta = 1/\Gamma$. }
}
\label{eLedust}
\end{center}
\end{figure}

{ Fig.\ \ref{eLedust} compares the apparent isotropic luminosities (or production spectra) of neutrons (blue curves) with neutrinos (red curves).  Using the same parameters as Fig.\ \ref{nLnangle}, this figure shows production spectra calculated for $\Gamma = 100, 50$ and 20 at observing angles $\theta = 0$ and $\theta = 1/\Gamma$. The comparison reveals the strong, $\delta_{\rm D}^5$ beaming factor of this process (Eq.\ \ref{gnln9}) when viewing off-axis, leading to a factor $2^5 = 32$ reduction of the production luminosity at the Doppler beaming angle compared to the luminosity along the jet axis. The relative fluxes as a function of $\Gamma$ at the same observing angle (which would be peculiar except for $\theta = 0$) is closer to $\propto \Gamma^4$, due to kinematic effects and normalization. 
Neutrons are made as secondaries of protons with $E_n \approx 0.8 E_p$ and carry the bulk of the energy of the incident protons. Neutrinos, by comparison, carry only $\approx 5$\% and $\beta$-neutrinos only $\sim 0.1$\% of the energy of the original neutron.  }


Figs.\ \ref{nLnangle} -- \ref{eLedust} apply to single one-zone blobs traveling through a beam production region of extent $R_{\rm BLR}=2\times{10}^{17}$~cm. The duration of the neutron production for on-axis observers 
is $\sim (1+z) R_{\rm BLR}/2\Gamma^2 c \sim 500$ s, and shorter if the accelerated proton distribution
decays before it traverses the length of the jet. 

{ Calculations of secondary neutron and neutrino production spectra from a continuous jet allow for larger secondary luminosities at the expense of variability. On the other hand, smaller values of $\Gamma$ might be more typical during these extended periods, and it is less likely that we would be looking directly along the jet axis. Even for smaller 
jet Lorentz factors $\Gamma \approx 20$, detection of neutron-induced or neutrino emission might be compensated by its longer duration, though compromised by the larger backgrounds for neutrino detection for the longer time windows.}

\section{Pair-Production Opacity and Photopion Efficiency for Isotropic Radiation Fields} \label{sec:opacity}

Our results to now are fairly intuitive. Cosmic-ray protons with energies $\gamma_p \gtrsim \bar\e_{\rm thr}/\e_\star$ interacting with an external radiation field with mean photon energy $m_e c^2\e_\star$ have a photohadronic neutron-production efficiency 
$\approx K_{p\gamma } \sigma_{p\gamma} R u_0/m_ec^2 \e_\star \sim 1$ -- 10\% in the { inner jet consisting of the BLR and IR dust torus radiation field local to the supermassive black hole in} 4C +21.35 (equation (\ref{etapgamma}) and Sections 3.2 and 3.3). 
With the stipulation that the absolute jet luminosity is Eddington-limited and that 
the bulk of this power is transformed to an $s=2$ spectrum of protons with maximum escaping energies $\approx 10^{20}$ eV, then an impulsive jet has to have $\dD\gtrsim 100$ to produce a highly collimated beam of outflowing UHE neutrons with apparent luminosity $\approx 10^{48}$ erg s$^{-1}$ 
in a pulse of particles $\lesssim {10}^{13}$~cm in width { measured in the black-hole frame}. 
The large Doppler factor collimates the beam and preserves the rapid variability. For a continuous jet, apparent isotropic neutron luminosities of $\approx 10^{46}$ erg s$^{-1}$ are formed by a jet with more modest Doppler factors. The outflowing neutron or neutrino fluxes will persist for longer times, though without short timescale variability.

We now turn our attention to the  $\gamma\gamma$ opacity and photopion efficiency of $\gamma$ rays and neutrons in the radiation field formed by the BLR and dust torus, for application to 4C +21.35 or other blazars. { We work under the assumption of local isotropy; the paper by \citet{1979A&A....76..306G} calculates energy density with this assumption relaxed.}

\subsection{$\gamma\gamma$ Opacity}

The opacity $\tau_{\g\g }(\e_1 )$ for a $\gamma$ ray with energy $m_ec^2\epsilon_1$  to travel a distance $R$ through an isotropic radiation field with energy density $u_0 = m_ec^2 \int_0^\infty d\e\,\e\, n_{\rm ph}(\e )$ consisting of photons with energy $m_ec^2\e$ described by the number density distribution $n_{\rm ph}(\e )$ is given through the relation
\begin{equation}
{d\tau_{\g\g}\over dx} = {\pi r_e^2 \over \e_1^2} \int_{1/\e_1}^\infty 
d\e\; \e^{-2} \; n_{\rm ph}(\e )\;\bar\varphi(s_0)\;, 
\label{dtauggdx_2}
\end{equation}
where $s_0 \equiv \e\e_1$,  
\begin{equation}
\bar\varphi(s_0 ) = 2\int_1^{s_0} ds 
\;{s\sigma_{\g\g}(s)\over \pi r_e^2}\;,
\label{barvarphi}
\end{equation}
and $ \sigma_{\g\g}(s)$ is the $\gamma\gamma$ pair-production cross section. Thus
\begin{equation}
\tau_{\gamma\gamma }(\e_1 )\cong {\pi r_e^2 R\over \e_1^2}\,\int_{1/\e_1 }^\infty d\e \,\e^{-2} n_{ph}(\e )\bar\varphi (s_0)\;.
\label{tgg}
\end{equation}
 The function 
$$\bar\varphi(s_0 ) = (2s_0 -2 +{1\over s_0})\ln w_0 + 2(1-2s_0)\sqrt{1-s_0^{-1}} + $$
\begin{equation}
\ln w_0 [4\ln (1+w_0) - 3 \ln w_0] - {1\over 3}\pi^2 
+ 4\sum_{n = 1}^\infty(-)^{n-1}n^{-2} w_0^{-n}\;,
\label{barvarphi_s0_6}
\end{equation}
where $w_0 \equiv 2s_0\sqrt{1-s_0^{-1}}-1$. The asymptotes of $\bar\varphi(s_0 )$ are 
\begin{equation} 
\bar\varphi(s_0 ) \rightarrow
 \cases{(2s_0+\ln 4s_0)(\ln 4s_0-2) 
- \cr  ~{\pi^2 -9\over 3} +s_0^{-1}(\ln 4s_0 +9/8)+ \dots\; &  $s_0 \gg 1$ \cr\cr
       {4\over 3}(s_0 - 1)^{3/2} + 
{6 \over 5} (s_0 - 1)^{5/2}-\cr  ~(253/70)(s_0 -1)^{7/2}+\dots\; 
	 &  $s_0 -1 \ll 1 $ \cr}\;
\label{barvarphi_s0}
\end{equation}
 \citep{gs67,bmg73}.

\begin{figure}[t]
\begin{center}
 \includegraphics[width=2.6in]{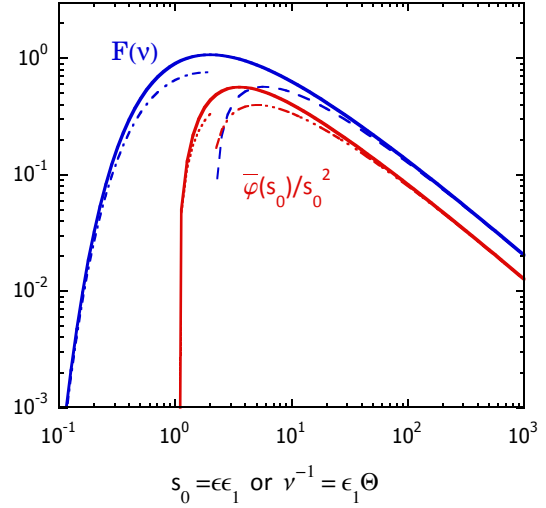}  
 \caption{The function $\bar\varphi(s)/s^2$, calculated from Equation (\ref{barvarphi_s0_6}), is given by the solid red curve, along with asymptotes in the low- and high-energy limits from Equation (\ref{barvarphi_s0}). The dark blue curve is the function ${\cal F}(\nu)$, Equation (\ref{calFnu}), with asymptotes given by  Equation (\ref{calFnu_asym}). }
\label{F(nu)}
\end{center}
\end{figure}

The pair-production opacity for a photon with energy $m_ec^2\e_1$ to pass through an isotropic monochromatic radiation field with target photon energy $m_ec^2 \e_\star$ is therefore
\begin{equation}
\tau_{\g\g}(\e_1) \cong \bar \tau_{\g\g}{\bar\varphi(s_0)\over s_0^2}\;,\; {\rm where }\; \bar \tau_{\g\g}  
\equiv {\pi r_e^2 R u_0\over m_ec^2 \e_\star} =\frac{3}{8} \frac{\sigma_T R u_0}{\epsilon_\star m_e c^2}\;.
\label{taugge1}
\end{equation} 
A graph of the function $\bar\varphi(s_0)/s_0^2$ used to calculate $\tau_{\g\g}(\e_1)$ in an isotropic photon field, including asymptotes from Equation (\ref{barvarphi_s0}) at small and large values of $s_0$, is shown in Fig.\ \ref{F(nu)}. Note the $\ln(\e_1)/\e_1$  dependence at $s_0 \gg 1$ because  $\bar\varphi(s_0)/s_0^2 \rightarrow 2\ln(0.54s_0)/s_0$ in the limit $s_0\gg 1$.  The function $\bar\varphi(s_0)/s_0^2$ reaches a maxiumum value of $\approx 0.56$ at $s_0 \cong 3.54$, and falls by a factor of $\approx 20$ when $s_0$ goes from 3.5 to 300. For Ly $\alpha$ line photons, the peak opacity corresponds to about 90 GeV in the stationary frame. Neglecting photon broadening processes, there is a sharp lower limit to the opacity at  $\e_1=1/\e_\star$, that is, at 25.6 GeV in the stationary frame for opacity from Ly $\alpha$ photons.

Substituting the graybody spectrum, Equation (\ref{ugbeTheta}), into Equation (\ref{dtauggdx_2}) gives
$$\tau^{\rm gb}_{\g\g}(\e_1) \cong {2\alpha_f^2 g R\over \lambda_{\rm C}}\;\e_1^{-2} \int_{1/\e_1}^\infty d\e\; {\bar\varphi(s_0)\over \exp(\e/\Theta)-1} $$
\begin{equation}
\equiv {2\alpha_f^2gR\over \lambda_{\rm C}}\;\Theta^3 {\cal F}(\nu)\; =\; {15 r_e^2 R\over \pi^3}\,{u_0\over m_ec^2 \Theta}\,{\cal F}(\nu)\;
\label{dtaubbggdx}
\end{equation}
for the energy-dependent $\g\g$ opacity through a graybody radiation field, where the fine-structure constant $\alpha_f = e^2/\hbar c \cong 1/137$, and 
\begin{equation}
\nu \equiv {1\over \e_1\Theta}\;.
\label{nudef}
\end{equation}
Here $g$ is the graybody factor representing the ratio $u_0/u_{\rm bb}$ of energy densities of the graybody and blackbody radiation fields with temperature $T$ (see Eq.\ \ref{ugbeTheta}), and the blackbody $\g\g$ opacity function
\begin{equation}
{\cal F}(\nu) \equiv \nu^2\int_\nu^\infty dw\;{\bar\varphi(w/\nu)\over \exp(w)-1}\;.
\label{calFnu}
\end{equation}
This function reaches its maximum value at ${\cal F}_{pk} \equiv {\cal F}(\nu_{pk}) \cong 1.076$ at $\nu_{pk} \cong 0.51$. The asymptotes of ${\cal F}(\nu )$ are 
\begin{equation}
{\cal F}(\nu) \;\rightarrow \;\cases{\sqrt{\pi\nu}\; 
\exp(-\nu)(1+{9\over4\nu} )\;, 
&$\nu \gg 1 {\rm ~ or~} \e_1\ll 1/\Theta$\cr\cr {\pi^2 \nu\over 3}\; 
	\ln(0.47/\nu ),&$\nu \ll 1 {\rm ~ or~}\e_1 \gg 1/\Theta$\cr}\;. 
\label{calFnu_asym}
\end{equation}
Fig.\ \ref{F(nu)} is a graph of the function ${\cal F}(\nu)$ along with its asymptotes.

\subsection{Photopion Efficiency}

The derivation of the photopion energy-loss rate of an ultrarelativistic neutron with energy $E_n = \g_nm_nc^2 = 10^{20} E_{20}$ eV in an external isotropic radiation field described by 
the function $n_{\rm ph}(\e)$ follows closely the derivation sketched in Section 3.3 for cosmic-ray protons in a jet, and is given by 
$$t_{\phi\gamma}^{-1}(\gamma_n) = 
{ c  \over 2\g_n^2 }\int_0^\infty d\e\;{n_{\rm ph}(\e )\over \e^2 } \int_0^{2\g_n\e} d\bar\e_r \,\bar\e_r \sigma_{\phi\gamma}(\bar\e_r ) K_{\phi\gamma}(\bar\e_r )$$
\begin{equation}
\approx { c \hat\sigma  }\int_{\bar\e_{\rm thr}/2\g_n}^\infty d\e\;{n_{\rm ph}(\e )} \left[1 - \left( {\bar\e_{\rm thr}\over 2\g_n\e }\right)^2 \right] 
\;, 
\label{tpgamma_1}
\end{equation}
using the approximation $\sigma_{\phi\gamma}(\bar\e_r) K_{\phi\gamma}(\bar\e_r )\cong \hat \sigma H(\bar\e_r-\bar\e_{\rm thr})$ where, as before, $\bar\e_{thr} \cong 400$ is the photopion threshold photon energy and $\hat \sigma = 70~\mu$b is the product of the photopion energy-loss proton cross section and inelasticity \citep{ad03}. In a graybody photon field given by Equation (\ref{ugbeTheta}) with graybody factor $g$, 
\begin{equation}
t_{\phi\gamma}^{-1}(\gamma_n) \cong 
{8\pi c g \hat \sigma \Theta^3\over \lambda_{\rm C}^3} \; I(\omega )\;,
\label{tpgamma_2}
\end{equation}
where 
$I(\omega ) \equiv  I_2(\omega ) - \omega^2 I_0(\omega )$, from Equation (\ref{I2x2I0x}). 
The parameter 
\begin{equation}
\omega = {\bar\e_{\rm thr}\over 2\g_n\Theta}\; 
\label{omega_0}
\end{equation}
characterizes the different regimes of photopion interaction of an ultrarelativistic cosmic-ray proton in a blackbody photon distribution, and whether the proton interacts with the exponentially declining number of photons in the Wien regime ($\omega\gg 1; \g_n\ll  200/\Theta$ or $E_n \ll 10^{18}$ eV$/(T_{\rm dust}/1200{\rm~K})$), or with the bulk of the distribution ($\omega \ll 1; \g_n\gg 200/\Theta; E_n \gg 10^{18}$ eV$/(T_{\rm dust}/1200{\rm~K})$).

\begin{figure}[t]
\begin{center}
 \includegraphics[width=3.0in]{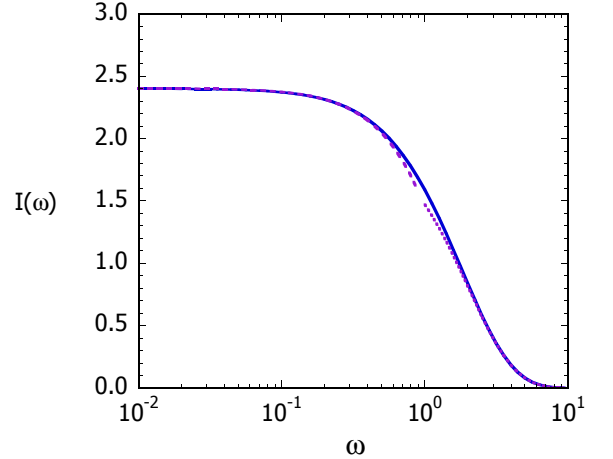} 
 \caption{The function  $I(\omega)$ and asymptotes, from Equation (\ref{I2x2I0x}), that describe the photopion energy loss rate of protons or neutrons with photons of a blackbody or graybody radiation field. }
\label{I(omega)}
\end{center}
\end{figure}

Expressing the photopion efficiency  
\begin{equation}
\eta_{\phi\gamma} = t_{\rm lc}/t_{\phi\gamma}\;,
\label{effphipi}
\end{equation} 
where the light-crossing  time scale $t_{\rm lc} = R/c$ across a region of size $R$, then
\begin{equation}
\eta^{\rm gb}_{\phi\gamma} (E_n) ={8\pi g \hat \sigma R \Theta^3\over \lambda_{\rm C}^3} \;I(\omega )\; =\;{15\over \pi^4}\,{\hat \sigma R u_0\over m_ec^2 \Theta}\,I(\omega )\;.
\label{etaphipigb}
\end{equation}
The photopion efficiency for monochromatic line radiation is simply
\begin{equation}
\eta^{\rm line}_{\phi\gamma} (E_n) ={u_0 \hat \sigma R \over m_ec^2 \e_\star}(1-x^{-2})H(x-1) \;,\; x \equiv {2\gamma_n\epsilon_\star\over \bar \epsilon_{\rm thr}} \;,
\label{etaphipiline}
\end{equation}
and approaches a constant value at $\g_n\gg \bar\epsilon_{thr}/2\e_\star$ (compare Equation (\ref{dEpoverdt})).
We show $\eta_{p \gamma}$ in Figure 9 using $R_{\rm BLR} = 2 \times {10}^{17}$~cm. One should keep in mind that $R=1$~pc should be used for the photomeson production between the neutron beam escaping from the BLR region and target photon fields from the dust torus, and beamed neutrons are essentially depleted in the dust torus. 

\subsection{Ratios}

For isotropic monochromatic  radiation, the ratio of the  peak $\g\g$  opacity to the photopion efficiency in the limit of high energy neutrons is 
$${\cal R}^{\rm line} = {\tau_{\gamma\gamma}(\e_1^{pk}) 
\over \eta_{\phi\pi } (x\gg 1)}= {\pi r_e^2\over \hat \sigma}\left[ {\bar\varphi(s_0)\over s_0^2}\right ]_{pk}$$
\begin{equation}
\cong 0.56\; {\pi r_e^2\over \hat \sigma} \cong 2000\;.
\label{Rlinephipi}
\end{equation}
The ratio of neutron energy at $x = 1$ and photon energy at the opacity peak is 
\begin{equation}
{\xi}^{\rm line}_{p\gamma} = {E_n(x = 1)\over E_1^{pk}} = {m_n\over m_e}\,{\e_{\rm thr}\over (2\times 3.54)} \simeq 10^5\;.
\label{Rlinepgamma}
\end{equation}

For thermal blackbody or graybody radiation, 
\begin{equation}
{\cal R}^{\rm thermal} =  {\pi r_e^2 \over \hat \sigma }{{\cal F}_{pk}\over I_{pk}} \cong 3600\;{1.076 \over \Gamma(3)\zeta(3)} \cong 1600\;,
\label{Rthermalphipi}
\end{equation}
and the ratio of neutron energy at $x = 1$ and photon energy at the opacity peak for the thermal photon distribution is
\begin{equation}
{\xi}^{\rm thermal}_{p\gamma} = {m_n\over 4 m_e}\,\e_{\rm thr} \simeq 2\times 10^5\;.
\label{Rlinepgamma1}
\end{equation}
Corresponding relations in relativistic jet sources are given by \cite{drl07}.


\subsection{Application to 4C +21.35} \label{sec:4C+2135}

\begin{figure}[t]
\begin{center}
 \includegraphics[width=3.0in]{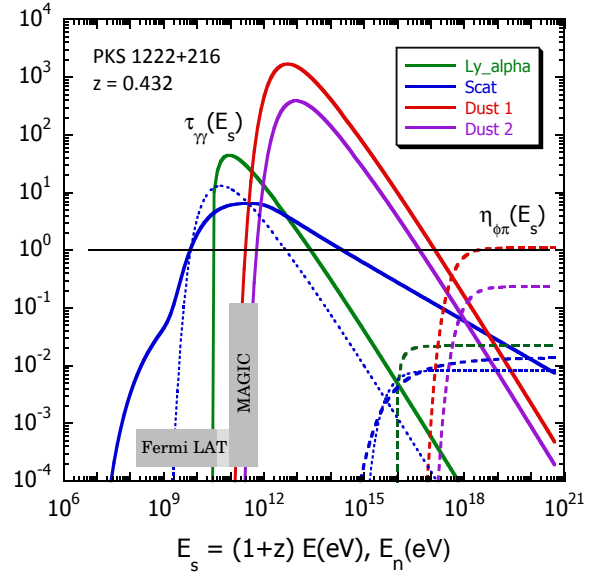} 
 \caption{Pair production opacities of $\gamma$ rays and photopion efficiencies of UHECR protons or neutrons in a monochromatic Ly $\alpha$ radiation field in the BLR, with the radiation field scattered by thermal electrons in the BLR, and with the dust radiation fields of the IR radiation field { found both in the BLR and} the dusty torus.  Photon energies $E({\rm eV})$ and neutron energies $E_n({\rm eV})$ are measured in the local source frame.
For $\gamma$ rays, $R=R_{\rm BLR}=2 \times {10}^{17}$~cm is used for the emission radius of the BLR region and $R=1$~pc is used for the emission radius of the dust torus. UHE neutron production is assumed to occur in the BLR region, { where both BLR and IR radiation fields are important}, and escaping neutrons are assumed to interact both with Ly $\alpha$ and scattered radiation within $R = R_{\rm BLR}$, but only with the IR radiation of the torus within $R = 1$ pc. Light dotted curves show opacities and photopion efficiencies using the graybody approximation for the scattered accretion-disk radiation field described by Equation (\ref{uacrdisk}).  
 }
\label{opaeff}
\end{center}
\end{figure}

From eqs.\ (\ref{uLyalphaTheta}) and (\ref{taugge1}), we find that the quasar BLR opacity to Ly $\alpha$ photons is 
\begin{equation}
\tau_{\g\g }(\e_1) \cong 40 R_{17}\phi_{-1}\,{\bar\varphi(s_0)\over s_0^2}
\label{taugge2}
\end{equation}
The Ly $\alpha$ opacity with $R_{17}\phi_{-1} = 2$ is shown in Figure.\ \ref{opaeff}. The large opacity in the VHE range precludes the photons detected with MAGIC from 4C +21.35 from being made in the BLR.

Using the \cite{mal11} parameters for the quasi-thermal IR emission from the hot dust, the graybody factor
\begin{equation}
g = {u_{\rm IR}\over u_{\rm bb}} = 0.18 {L_{46}\over {R_{\rm pc}}^2} {\left( \frac{T}{1200~\rm K} \right)}^{-4}\;,
\label{gdust}
\end{equation}
taking $u_{\rm IR} \cong L/4\pi R^2 c$, and noting that the energy density $u_{\rm bb}(1200$ K$) = 0.0156$ erg cm$^{-3}$ of a 1200 K blackbody radiation field. From Equation (\ref{dtaubbggdx}),
\begin{equation}
\tau^{\rm bb}_{\g\g}(\e_1) = {2\alpha_f^2 gR\over \lambda_{\rm C}}\;\Theta^3 {\cal F}(\nu)\;\cong \; 2000\,{L_{46}\over R_{\rm pc}} {\cal F}(\nu)\;,
\label{dtaubbggdx1}
\end{equation}
with a peak opacity at $\nu \cong 0.5$, implying $E_1^{pk} \cong 5$ TeV. For the warm dust component at $T = 660$ K,
\begin{equation}
\tau^{\rm bb}_{\g\g}(\e_1) \;\cong \; 350\,{L_{45}\over R_{\rm pc}} {\cal F}(\nu)\;,
\label{dtaubbggdx2}
\end{equation}
with a peak opacity at $E_1^{pk} \cong 10$ TeV.

In addition, there is the scattered radiation field from  free electrons in the BLR that scatter radiation emitted by the optically thick Shakura-Sunyaev accretion disk and a hypothetical hot X-ray pair plasma found close to the black hole. The optical depths to these photon fields are calculated using Equation (\ref{tgg}) and the relation 
\begin{equation}
m_ec^3\e^2 n_{\rm ph}(\e) = \e u_{sc}(\e ) \approx \tau_{\rm sc}\,{\e L(\e )\over 4\pi R^2 c}\;
\label{eusce}
\end{equation}
for the scattered radiation field. The Shakura-Sunyaev disk spectrum in the form
\begin{equation}
\e L_{\rm disk}(\e ) = {L_{\rm disk}\over \Gamma(4/3)}\left( {\e\over \e_{\rm max}}\right )^{4/3}\exp(-\e/\e_{\rm max}) \;
\label{eldisk}
\end{equation}
is used to estimate $\tau_{\g\g}(\e_1)$ from its scattered radiation field. For a hot plasma making an X-ray spectrum with index ($\alpha = 1/2$) and luminosity $L_X$, 
\begin{equation}
\e L^{X}(\e ) \cong {L_{X}\over \Gamma(1/2)}\left( {\e\over \e_{\rm max}}\right )^{1/2}\exp(-\e/\e_{\rm max})\;H(\e - \e_{\rm min}) \;.
\label{elX}
\end{equation}
The normalization becomes increasingly accurate when $\e_{\rm min}\ll\e_{\rm max}$.

Fig.\ \ref{opaeff} shows calculations for a disk spectrum with luminosity $L_{\rm disk}= 2\times 10^{46}$ erg s$^{-1}$ and $E_{\rm max,disk} = 50$ eV, and an X-ray corona with $L_{X}= 4.5\times 10^{45}$ erg s$^{-1}$ and $E_{{\rm max},X} \cong 5$ keV, using a dust covering factor of  22\% \citep{mal11} and taking the scattering optical depth $\tau_{\rm sc} = 0.01$. The photopion efficiency was determined for the scattered radiation fields by substituting Equations\ (\ref{eldisk}) and (\ref{elX}) into Equation (\ref{tpgamma_1}), using Equation (\ref{eusce}) and Equation (\ref{effphipi}). For comparison, the dotted curves show the opacity and photopion efficiency when the scattered accretion-disk radiation field is approximated by a thermal graybody spectrum using the parameters from Equation (\ref{uacrdisk}). Note that  radiation fields with higher effective temperatures but comparable luminosities provide negligible additional opacity because of the smaller density of target photons.

\section{Synchrotron Radiation from UHECR Secondaries} \label{sec:sync}

\subsection{Analytical considerations}

The results in Figure\ \ref{opaeff} show that the multi-GeV $\gamma$ rays detected from 4C +21.35 with MAGIC cannot be made in the BLR, and that inner jet radiation should be strongly attenuated above several GeV from scattered accretion-disk radiation. For an accretion-disk power of $5\times 10^{45}$ erg s$^{-1}$, $R_{17}\approx 2$ from Equation (\ref{RBLR}), and if the accretion-disk power is even larger, as indicated by the Swift data \citep{tav11}, the $\gamma\gamma$ attenuation is likely to be even larger than shown in Fig.\ \ref{opaeff}. Thus the emission region for VHE photons must at least be situated beyond the edge of the BLR, but even then photons with energies $\gtrsim 200$ GeV are strongly absorbed by the IR radiation from the torus. The actual attenuation depends precisely on deviations of the dust spectrum from a graybody in the Wien regime, but even with the exponentially declining number of photons in the near infrared spectrum, there is still large opacity for VHE emission above a few hundred GeV. The lack of attenuation in the MAGIC data to $\sim 400$ GeV ($\sim 600$ GeV in the source frame) implies that the VHE $\g$ rays have to be made at and beyond the pc scale.

Escaping neutrons easily leave the BLR but suffer strong photopion losses when passing through the infrared field of the dust torus, where neutrons would deposit a large fraction of their energy during escape. 
Hence, we should expect formation of beams consisting of UHE $\gamma$ rays, pairs and neutrinos.
The photopion production efficiency $\eta_{\phi\gamma}(\g_n)$ of an escaping neutron or proton with Lorentz factor $\g_n$ passing through, in particular, the hot dust radiation field of 4C +21.35, 
is estimated to be $\eta_{\phi\gamma}(\g_n) \sim 2.5 {(u_{0}/\epsilon_\star)}_4 R_{\rm pc}$ 
(see Equation (\ref{etapgamma}) and  Fig.\ \ref{opaeff}, noting $\langle \e_\star \rangle \cong 2.7 \Theta$), and this energy is transformed into ultra-relativistic e$^+$-e$^-$ pairs, photons and neutrinos.

In the case of 4C +21.35, efficient photopion losses apply to neutrons or protons with $\omega \ll 1$, which for hot dust with $\Theta\cong 2\times 10^{-7}$ implies a range of energies
\begin{equation}
m_pc^2 \; {\e_{\rm thr}\over 2\Theta} \cong 6\times 10^{17} {\rm~eV} \lesssim E_p\lesssim E_{\rm max} \equiv 10^{20}E_{20}{\rm~eV} \;
\label{EeV}
\end{equation}
for which photopion losses are large, as can be seen from Figure\ \ref{opaeff}. 

In each  $n+\gamma\rightarrow p +\pi^-$ interaction, roughly 20\% of the neutron's original energy is given to two pionic photons, $\approx 5$\% is given to an electron, and $\approx 15$\% is given to three neutrinos. Secondary electron energies therefore range in value from $\approx 3\times 10^{16}$ eV to $2\times 10^{18}E_{20}$ eV, implying Lorentz factors in the range
\begin{equation}
6\times 10^{10} \lesssim \g_{e} \equiv 10^{11}\g_{11}\lesssim 10^{13} E_{20},
\label{gammae}
\end{equation}
so $0.6 \lesssim \g_{11}\lesssim 100E_{20}$.
The synchrotron photons from these hyper-relativistic\footnote{These electrons are ``hyper-relativistic" in the sense \citep{da04} that such high-energy electrons cannot be directly accelerated by traditional Fermi acceleration mechanisms that compete directly with synchrotron losses.} be produced through  secondary e$^+$ and $e^-$ have observed energies
\begin{equation}
E_{\rm syn} = m_ec^2 \e_{\rm syn} \cong  {3m_ec^2B \gamma^2\over 2 B_{\rm cr}(1+z)} \cong 120 {B_{\rm \mu G,pc}}\g_{11}^2\;{\rm MeV}\;,
\label{esyn}
\end{equation}
where $B_{\rm cr} = 4.414\times 10^{13}$ G is the critical magnetic field and $B_{\rm pc}$ is the magnetic field at the pc scale. Thus the photon energies of the first generation synchrotron spectrum are in the range
\begin{equation}
43 {\rm~MeV} \lesssim {E_{\rm syn}\over B_{\rm \mu G, pc}} \lesssim 1.2 E_{20}^2\;{\rm TeV}\;.
\label{Esyn}
\end{equation}
The second-generation synchrotron spectrum from e$^+$-e$^-$ pairs made when synchrotron $\gamma$ rays form pairs through $\g\g$ attenuation is important for larger, $\sim$ mG fields. If a synchrotron $\gamma$ ray forms a pair, with each electron and positron receiving half the energy,\footnote{This is not a good approximation when the photons pair-produce in the Klein-Nishina (KN) portion of the pair-production cross section \citep[see, e.g.,][]{mur09,kal11}. Escape of UHE $\gamma$ rays depends on the internal target photon spectrum, which may be suppressed by self-absorption of the synchrotron spectrum or the Rayleigh-Jeans portion of the black-body spectrum \citep{rmz04,mur09}.} then the second-generation synchrotron spectrum ranges between 
\begin{equation}
0.052 {\rm~eV} \lesssim {E_{\rm syn}\over B_{\rm mG,pc}^3} \lesssim 35 E_{20}^2\;{\rm MeV}\;.
\label{Esyn2}
\end{equation}
However, those second generation synchrotron photons are typically irrelevant for the highly variable emission (see below). 

\begin{figure}[t]
\begin{center}
 \includegraphics[width=3.5in]{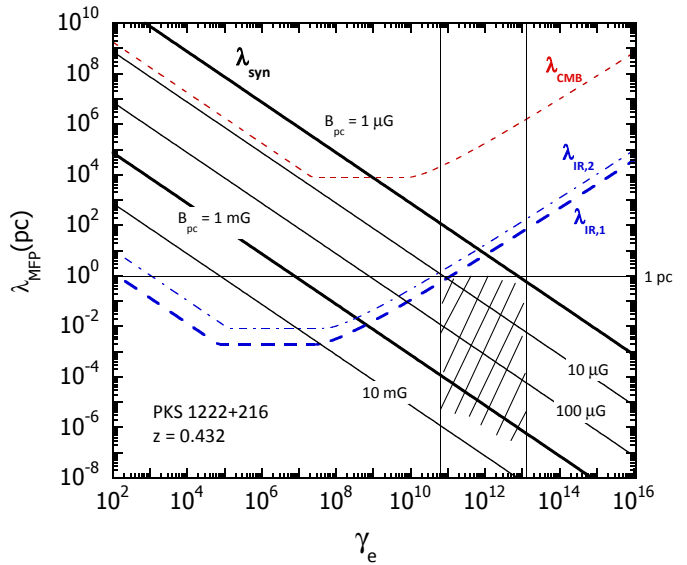} 
 \caption{Energy-loss mean-free-path (MFP) for synchrotron and Compton losses of an electron with Lorentz factor $\g_e$. Synchrotron loss MFPs are calculated in a tangled magnetic field ranging from 1 $\mu$G to 10 mG, as labeled. Compton energy-loss MFPs are plotted for the CMB radiation, and for graybody IR radiation fields from hot ($T = 1200$ K) and warm ($T = 660$ K) dust, with a graybody factor corresponding to a distance $R = 1$ pc and properties given by \citet{mal11}. Vertical lines represent the range of injection electron Lorentz factors for electrons formed by the decay of $\pi^+$ from photopion interactions of UHECR neutrons.The cross-hatched region represents the range of $\g_e$ and magnetic fields where 10 minute variability can be preserved (see text). }
\label{MFPfig}
\end{center}
\end{figure}

Synchrotron losses will dominate when the synchrotron energy-loss MFP is shorter than other energy-loss MFPs. Fig.\ \ref{MFPfig} shows the synchrotron energy-loss MFP in comparison with the Compton energy-loss MFP on the dust radiation fields and, for comparison, the CMBR field at $z = 0.432$. The electron synchrotron energy-loss rate assumes a randomly oriented magnetic field on the pc scale of strength $B_{\rm \mu G,pc}$. The Compton energy-loss MFP of an electron passing through a graybody radiation field is approximated in Fig.\ \ref{MFPfig} by the expression
$$\lambda_{\rm C}(\g )\equiv c {\left |{\dot \g_{\rm C}\over \g} \right|}^{-1}=$$
\begin{equation} 
\cases{  {45\lambda_{\rm C}^3\over 32\pi^5 g\sigma_{\rm T} \Theta^4  \g}
\cong {106 R_{pc}^2\over L_{46} \g}\,{\rm pc},&  $\g \leq \bar\gamma 
\equiv {0.015\over \Theta} 
\cong {7.4\times 10^4\over T_{1200}}$ \cr\cr 
      {1.52\times 10^{-3} R_{pc}^2T_{1200}\over L_{46}}\,{\rm pc}, &  $\bar\g < \g\leq \g_\star \equiv
{4.92\over\Theta} $ \cr\cr  
{2\lambda_{\rm C}^3\g\over  \pi^3 g\sigma_{\rm T}\Theta^2\ln (0.55\g\Theta )}\;\cong\;
\cr ~~
{6.2 \g_{11}T_{1200}^2 R^2_{pc}\over L_{46}\ln (1.1\times 10^4 \g_{11}T_{1200})}\;{\rm pc},\;
	 &  $\g > \g_\star \cong 
{2.4\times 10^7\over T_{1200}}$ \cr}
\label{gammadotgb}
\end{equation}
where the energy density of the blackbody radiation field with temperature $\Theta$ is given by 
Equation (\ref{ubbTheta}), and $T_{1200}\equiv T/1200$ K. This simple expression connects the Thomson and extreme KN asymptotes with a constant MFP in the intermediate regime determined by equating the Thomson MFP with the value of the KN MFP evaluated at the electron Lorentz factor $\g_\star$ given by $\ln (0.55\g_\star\Theta ) = 1$. 

From Figure\ \ref{MFPfig}, one sees that synchrotron losses of electrons formed by neutron photopion production dominate Compton losses when $B_{\rm pc}\gtrsim 3 \mu$G, and Equation (\ref{Esyn}) shows that first-generation synchrotron photons with energies exceeding a few GeV are formed when $B_{\rm pc}\gtrsim 10 \mu$G. 
We now show that such a process can accommodate the short variability timescale observed with MAGIC if the inner engine generates pulses of UHE neutrons that are modulated by activity of the inner jet on short timescales. In this case, synchrotron production at distance $R$ from the central engine can vary on timescale $\Delta t$ if the emission is made within the angle $\theta_t$ of the line of sight according to the relation $(1+z)R(1-\cos\theta_t )/c = \Delta t$. This is satisfied for electrons produced within the angle 
\beq
\theta_{\rm dfl} \lesssim \theta_t = \sqrt{2c\Delta t \over R(1+z)} \cong 2.9\times 10^{-3} \sqrt{(\Delta t/600{\rm~s})\over (1+z) R_{\rm pc}}\;.
\label{thetat}
\eeq
 The electrons making the GeV/TeV synchrotron radiation are deflected by the angle $\theta_{\rm dfl} \cong \sqrt{2/3} \lambda_{\rm syn}/r_{\rm L} = 6 \sqrt{2/3} \pi e/\sigma_{\rm T} B\gamma^2$, 
where $\lambda_{\rm syn} = 6\pi m_ec^2/\sigma_{\rm T} B^2\gamma$ is the 
energy-loss length for synchrotron losses and $r_{\rm L} = m_ec^2\gamma/eB$ is the electron Larmor radius.
The criterion that $\theta_{\rm dfl}< \theta_t$ \citep{mur12} implies 
\beq 
B_{\rm \mu G,pc}\gamma_{11}^2 \gtrsim 330 \sqrt{({\Delta t}/600~{\rm s}) \over R_{\rm pc}}
\label{Bgamma2}
\eeq
to preserve observed variability on a timescale shorter than $\approx 600$ s for 4C +21.35.
Together with Equations (\ref{Esyn}) and (\ref{Esyn2}), this variable synchrotron radiation will be detected 
at photon energies 
\beq
40 \sqrt{({\Delta t}/600~{\rm s}) \over R_{\rm pc}} \lesssim E_{\rm syn}({\rm GeV}) \lesssim 1.2\times 10^3 E_{20}^2 B_{\rm \mu G,pc} \;.
\label{EsynGeV}
\eeq
The constraints will be relaxed if the coherence length of the magnetic field is small in comparison with the synchrotron cooling length. 

A number of important points can be made regarding this result. The synchrotron radiation frequency depends only the product $B\gamma^2$ and not on $B$ or $\gamma$ separately. 
Hence, $\Delta t$ essentially depends on $E_{\rm syn}$, without the explicit dependence on $B$ and $\gamma$ \citep{mur12}. Thus highly variable synchrotron emission is expected from VHE synchrotron $\gamma$ rays.  
Because the photohadronic production kinematics limits the lower bound of the secondary electron distribution to be at $\gamma_{11}\approx 0.6$, from Equation (\ref{gammae}) and Figure \ref{opaeff}, the characteristic magnetic field at the pc scale is implied to be $\lesssim 3$ mG, otherwise the energy range of synchrotron photons is out of the MAGIC band.  Lower magnetic fields are allowed as long as the synchrotron cooling length is short enough, where lower $B$ corresponds to higher $\gamma$ at given $E_{\rm syn}$.  A definite prediction of this model is that the variability times of the $\gamma$-ray synchrotron radiation become longer at lower energies, because a larger angular range of cooling and deflecting electrons can then contribute to the observed emission. Emission from the inner jet can, however, dominate  at GeV energies, so this prediction applies only to the VHE synchrotron $\gamma$ rays.
Inverse-Compton cascade radiation should be more slowly variable because $\lambda_{\rm IR} > \lambda_{\rm syn}$, and second-generation synchrotron emission (Equation (\ref{Esyn2})) does not contribute to the highly variable emission since it is typically expected at lower energies.

\subsection{Numerical Results}

To demonstrate the resulting secondary spectra produced by the neutron beam launched from the BLR region, we also perform numerical calculations. In order to evaluate photon and pair yields from the photomeson production, we use SOPHIA \citep{muc+00} and solve the kinetic equations for injected photons and pairs,  
as in \cite{mdtm11}. 
We calculate cascades taking into account synchrotron and inverse-Compton emissions, and we focus on beamed emissions such that $\theta_{\rm dfl} < \theta_{t}$ to evaluate variable emission components produced in the dust torus. In this work, we only show the highly variable emission component with $\Delta t < 600$~s, and we do not include emissions 
with longer timescales, though they, as well as radiation from the inner jet, partially contribute to the overall received flux \citep{mur12}.   

Figure \ref{magic1} 
show results for the case where the neutron beam is produced by photomeson interactions between UHE protons and external photons in the BLR.  The spectrum used here is the same as { that shown in  the inner jet calculation}  
of Figure \ref{nLnangle}, { which includes target Ly $\alpha$, scattered accretion-disk, and cool and warm torus radiation as target radiation fields.} 
Then, we calculate the photomeson production by UHECR neutrons in the external photon fields of the dust torus, whose radius is set to $R_{\rm dust}=1$~pc.  The magnetic field is set to 10~$\mu$G.  In the calculations, the dust temperatures are set, as before, to $T_{\rm dust1}=1200$~K and $T_{\rm dust2}=660$~K, with luminosities $L_{\rm dust1}=8 \times {10}^{45}~{\rm erg}~{\rm s}^{-1}$ and $L_{\rm dust2}={10}^{45}~{\rm erg}~{\rm s}^{-1}$, 
respectively.\footnote{For numerical calculations, we use $\tau_{\rm sc} L_{\rm disk}=5 \times {10}^{44}$ erg s$^{-1}$ and $\tau_{\rm sc} L_{X}=4.5 \times {10}^{44}$ erg s$^{-1}$, respectively. The EBL, CMB and cosmic radio background are also included, though they are irrelevant for source cascades.}   

We found that the required absolute jet power is $\sim {10}^{46}~{\rm erg}~{\rm s}^{-1}$, which is consistent with analytical expectations.  
From Figure \ref{opaeff},  one expects that UHE photons might escape from their emission regions, where they cascade in larger-scale magnetized regions, including elliptical galaxies, galactic winds, galaxy clusters and filaments \citep{mur12}.  
Since strong radio emission is presumably produced by kpc jets in FSRQs, UHE $\gamma$ rays can also be quickly depleted by such an additional target radio field, whose scale and intensity is uncertain.  For demonstration, we therefore consider an additional radio field which would be provided by the kpc jet.  From \citet{tav10}, the radio field is optimistically assumed to have a typical size of $R_{\rm radio} = {10}^{21}$~cm and a characteristic $\nu L_{\nu}$ radio luminosity $ = {10}^{43}~{\rm erg}~{\rm s}^{-1}~(\nu/{10}^{9}~{\rm Hz})$ at $\nu< {10}^{11}$~Hz, with a naive extrapolation to lower frequencies. 
 
To calculate this contribution, we also evaluate cascades initiated by UHE photons leaving the dust torus, using $c\Delta t \approx (1+z) (1- \cos \theta_{\rm dfl}) {\rm min}[\lambda_{\gamma \gamma}, R_{\rm radio}]$. Rapidly variable emission is still possible for the UHE photon-induced emission as long as $\lambda_{\gamma \gamma}$ is short enough \citep{mur12}.  The steady VHE cascade emission induced by the kpc scale jet is predicted, from  Fig.\ \ref{magic1},  to be at a lower flux level than the rapidly variable VHE emission formed at the pc scale. { The search for the low-level, high-energy plateau emission is feasible with the {\it Cherenkov Telescope Array} (CTA)  \citep{cta10} in development. Besides the UHE photons and leptons generated by neutron photomeson processes at the pc scale (thin solid curves), the subsequent cascade $\gamma$-ray spectrum formed in 
the dust torus is shown by the dashed curve. The UHE photons escaping into the kpc jet can experience further cascades, which makes an additional high-energy component with rapid variability. 

\begin{figure}[t]
\begin{center}
 \includegraphics[width=3.0in]{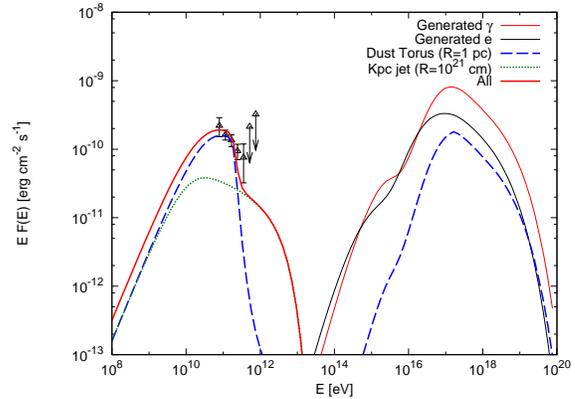} 
\caption{$\g$-ray spectra formed by a neutron beam generated in the inner jet, based on the calculations shown in Figure\ \ref{nLnangle}. MAGIC data are also overlaid, which have been deabsorbed using a low-intensity model of the EBL \citep{2011MNRAS.410.2556D}.}
\label{magic1}
\end{center}
\end{figure}

\begin{figure}[t]
\begin{center}
 \includegraphics[width=3.0in]{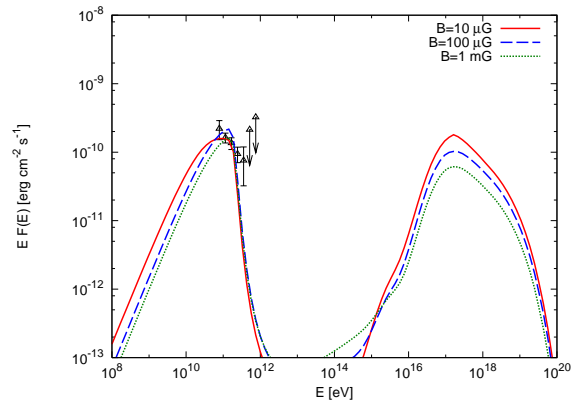} 
\caption{$\g$-ray spectra formed by a neutron beam generated in the inner jet, based on the calculations shown in Figure\ \ref{nLnangle}, but for different values of $B$.  The required absolute jet powers are $9.5 \times {10}^{45}~{\rm erg}~{\rm s}^{-1}$, $6.8 \times {10}^{45}~{\rm erg}~{\rm s}^{-1}$ and $4.1 \times {10}^{45}~{\rm erg}~{\rm s}^{-1}$ (top to bottom). }
\label{magic2}
\end{center}
\end{figure}

\begin{figure}[t]
\begin{center}
 \includegraphics[width=3.0in]{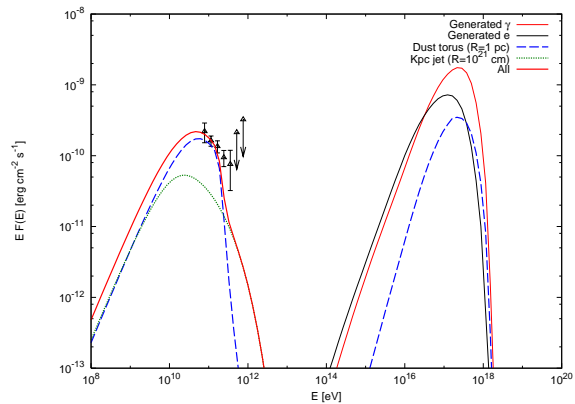} 
\caption{$\g$-ray spectra formed by a neutron beam generated in the inner jet, based on a power-law neutron spectrum with $s=0.8$ at ${10}^{15}~{\rm eV} < E_n < {10}^{18.5}~\rm eV$.  We use $\int d E_n \, L_n(E_n) = {10}^{49}~{\rm erg}~{\rm s}^{-1}$.}
\label{magic3}
\end{center}
\end{figure}

The results are not very sensitive to $B$ as long as the magnetic field is in the 
range $10~\mu {\rm G} \lesssim B \lesssim {\rm mG}$, as shown 
in Figure \ref{magic2}.  
Just for comparison, 
in Figure \ref{magic3}
we also show the case where the neutron beam is produced by photomeson production between UHE protons and nonthermal photons that can be generated by electrons accelerated at internal shocks. Here, we inject beamed neutrons with a power-law spectrum, 
 $N_n (E_n) \propto E_n^{-0.8}$. Such a hard spectrum can be realized 
if the target photon spectrum in inner jets has $\beta = 2.2$ and the 
proton spectrum has $s=2.$0 (App.\ B). 
The target photon spectrum is taken not to contradict 
 the observations. 
As in Fig.\ \ref{magic2}, the dashed curve represents the contribution 
of cascade gamma rays developed in the dust torus, while the 
dotted curve is for the additional kpc jet component. 
In this case, however, considerably larger absolute powers 
are required in order to explain the observations.}

\begin{figure}[t]
\begin{center}
 \includegraphics[width=3.0in]{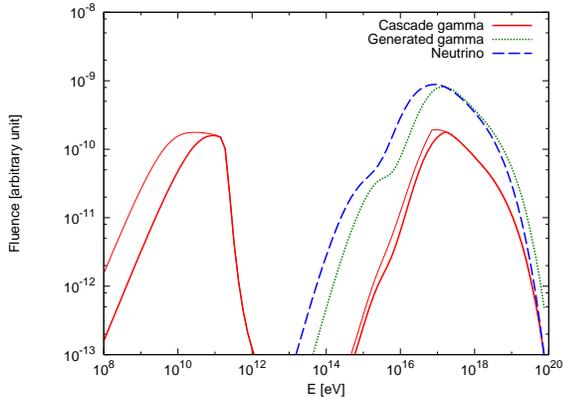} 
 \caption{Neutrino and cascade $\gamma$-ray spectra formed by a neutron beam generated in the inner jet, based on the calculations shown in Figure\ \ref{nLnangle}. Thick lines are for $\theta_{\rm dfl} < \theta_t$, while thin lines are for $\theta_{\rm dfl} < \theta_j=0.01$. { Solid curves show the cascade $\gamma$-ray spectrum 
formed in the dust torus, where the kpc jet component is not included.
Note that $\gamma$ rays with $\theta_{dfl} > \theta_t$ have longer time scales, and 
the flux is suppressed at lower energies.}  }
\label{magic4}
\end{center}
\end{figure}

In Figure \ref{magic4}, neutrino spectra are shown. Without performing a detailed calculation of neutrino detectability using the IceCube detector response, estimates show that a fluence of $\approx 10^{-4}$ -- $10^{-3}$ erg cm$^{-2}$ in $\gamma$ rays associated with photopion neutrino production at PeV energies is required for detection of a few PeV neutrinos with IceCube \citep{ad01}. 
Since the differential high-energy neutrino flux of $\approx 10^{-9}$ erg cm$^{-2}$ s$^{-1}$ is 
associated with photopion losses, then flaring episodes of days to weeks would be required to detect neutrinos with a km-scale neutrino telescope.  Although the expected fluence is not enough to expect neutrino detection by IceCube during a single flaring event from 4C +21.35, stacked flaring episodes of many such events, or searches for PeV neutrinos from $\gamma$-ray bright FSRQs like 3C 454.3, 3C 279, or PKS 1510-089 with high-energy neutrino telescopes are crucial for testing the model.  

\section{Discussion and Summary} \label{sec:discussion}

{ The sources of the UHECRs are unknown, but acceleration in the inner jets of blazars has been widely considered as a possible solution to this problem \citep{mb92,bgg06,der09,mt09}. In this paper, we argue that the detection of VHE radiation from the FSRQ 4C +21.35 supports this scenario, and provides a solution to the rapid VHE  variability and large $\gamma$-ray luminosity as an effect of ultra-high energy proton accelerated in the inner jet.\footnote{FSRQ blazars are scarce within the GZK volume, and their average emissivity is much less than blazar BL Lac objects, which may therefore be the preferred blazar class to accelerate most of the highest energy cosmic rays \citep{dr10,mdtm11}.} } 

A new feature of this study is the derivation of beaming factors for secondaries formed by photohadronic processes. Synchrotron and SSC emissions, and radiations from photohadronic processes with internal isotropic radiation fields, have a beaming factor $\propto\delta_{\rm D}^4$ due to compression in solid angle and time and enhancement in energy. External Compton scattering has a $\delta_{\rm D}^6$ beaming factor because Compton scattering is proportional to target photon energy density, which is boosted by two powers of $\dD$ when transformed to the comoving frame. Photohadronic processes with external isotropic photon sources have a beaming factor $\propto \delta_{\rm D}^5$, because photohadronic production is proportional to the target photon number density, which is increased by a single power of the Dopper factor in the comoving frame. Threshold and spectral effects complicate the beaming factors, and the more detailed relations,  including a formalism to calculate production spectra of photohadronic secondaries, are presented in this paper.

In Appendix \ref{app:A}, we also derived accurate photon powers for bolometric $\gamma$-ray fluxes associated with synchrotron and SSC processes, and for external Compton processes. These relations demonstrate the much stronger decline of flux for $\gamma$ rays from external Compton scattering compared to SSC fluxes, which may help explain the relative number of misaligned radio galaxies of different types \citep{2010ApJ...720..912A}.

The main goal of this paper is to propose a new mechanism to generate rapidly variable $\gamma$ rays at large distances from a black-hole engine in order to explain the puzzling observations of 4C +21.35. { The variability timescale is determined by processes in the inner jet that relate to the dynamical timescale of the black hole. For black hole masses of 1.5 -- 8 $\times 10^8 M_\odot$ \citep{2004ApJ...615L...9W,2012ApJ...748...49S}, the dynamical timescale $\approx 1$ -- $10 \times$ larger than the variability timescale. Some of the same solutions applied to rapid variability in TeV blazars,  \citep[e.g.][]{bfr08,2012MNRAS.420..604N}, could operate in the inner jet of 4C +21.35 if the larger black-hole mass is correct. }

We assume that the inner jets of blazars accelerate UHECRs that undergo photohadronic losses with ambient radiation fields and generate UHE neutrons, neutrinos, and $\gamma$ rays \citep{ad03}. Outflowing UHECR neutrons undergo photohadronic losses with IR torus photons to make pions that decay into hyper-relativistic leptons. If the magnetic field is $\sim 0.01-1$~mG, as shown by the cross-hatched region in Figure\ \ref{MFPfig}, then the VHE synchrotron radiation preserves the rapid variability of the inner engine at the pc scale. This mechanism is proposed as the origin of the rapid variability of 70 -- 400 GeV $\gamma$ rays observed with the MAGIC telescopes.   Only a very narrow angular range of electrons will contribute to the observed VHE radiation, so that the variability of the central source is preserved even if the opening angle of the relativistic jet is not narrow.  UHE photons escaping to larger distances, e.g., at the Mpc-scale region in the large scale structure surrounding the source may furthermore produce a slowly variable VHE synchrotron pair echo with timescales of $\sim 0.1-1$~yr \citep{mur12}, while UHE protons escaping to the magnetized regions can give almost non-variable signals \citep{ga05,kal11}. 
These radiations are useful as characteristic signals of UHECR acceleration, and are potentially observable with the planned CTA \citep{cta10}.
  
The $\gamma$-ray fluxes made principally by the synchrotron mechanism become less variable with decreasing energy, but at sufficiently low energies, $\gamma$ rays from the inner jet can avoid $\gamma\gamma$ absorption and additionally contribute to the observed flux. Indeed, Figure \ref{opaeff} shows that the $\gamma\gamma$ absorption for inner jet $\gamma$ rays becomes significant only above a few GeV. This corresponds precisely to the range of energies where the Fermi-LAT has discovered spectral breaks in FSRQs and low- and intermediate synchrotron peaked blazars \citep{2009ApJ...699..817A,2010ApJ...710.1271A,2010ApJ...721.1425A}. Whether this is a coincidence, or a cause of the GeV spectral breaks, it is pertinent that most models for the GeV breaks invoke production within the BLR \citep{ps10,fd10,2010ApJ...721.1383A,sp11}. An inner jet origin of the GeV radiation is consistent with the model proposed here, and the synchrotron $\gamma$ rays at multi-GeV/TeV energies would appear as a separate spectral component emerging from the attenuated inner jet radiation spectrum, which furthermore display increased variability with energy.

An accurate fit to the combined Fermi and MAGIC $\gamma$-ray spectrum requires detailed modeling of the radiation in the inner jet in addition to the subsequent cascade radiations formed as the VHE synchrotron photons pass through the IR photon field. Besides the leptonic emissions in the inner jet, a model involving the UHECR protons that are accelerated and undergo photopion losses to make escaping UHECR neutrons must be included.
 As noted earlier, long-term average apparent $\gamma$-ray luminosities $\gtrsim 10^{48}$ erg s$^{-1}$ are common in FSRQs, so apparent jet powers must even be larger. The apparent luminosity in UHECRs accelerated in the inner jet of 4C +21.35 must be $\approx 10^{48}$ -- 10$^{49}$ erg s$^{-1}$ to compensate for the inefficiencies to produce neutrons in the inner jet and pions at the pc scale.  For the target photons from the external BLR and IR torus fields, the radiative efficiency for $p\gamma$ photopion processes and the requirement of rapid variability from a single blob leads to the requirement of large Doppler factors, $\sim 100$ in order to make an apparent luminosity $\approx 10^{48}$ erg s$^{-1}$ in outflowing neutrons from a jet whose absolute power is Eddington-limited. 

Internal shocks formed by colliding plasma shells ejected from the nuclear black hole are often considered as the mechanism that dissipates energy into broadband nonthermal radiation. The constraint $R \lesssim \Gamma \dD c t_{\rm var}/(1+z)$ on variable emission from colliding shells means that $R\lesssim 0.01 (\Gamma/50)^2 (t_{\rm var}/600{\rm~s})$ pc, for $\Gamma \approx \dD$ \citep[e.g.,][]{tav10}. If the location of the emission site is required to be at a distance larger than a few pc from the central black hole, then unless $\Gamma \gtrsim 10^3$, which exceeds by an order of magnitude the outflow Lorentz factor inferred from any blazar data, such a model for the MAGIC observations cannot explain the VHE radiation due to the severe attenuation. 

The necessity to reconcile the contrary features of short term variability, which points to an inner jet origin, and detection of VHE $\gamma$ rays that must originate at the pc scale, has led to several proposed solutions.  \citet{tav11} consider a system with a compact zone outside the BLR and a second, possibly more extended zone either inside or outside the BLR to reproduce the complete SED. The compact emitting regions making the VHE radiation must have large Doppler factors, $\dD\approx 75$. \citet{nal12} consider various energetic constraints on 4C +21.35 system, and  also arrive at the requirement of compact emitting zones at pc scales.  The question then is the origin of these compact regions. One possibility is the formation of compact recollimation shocks \citep{mar06,bl09} that are produced when the external medium pressure overcomes the radiative jet, as proposed for M87 flares or the radio cores of blazars. The large powers  from 4C +21.35 challenge such an origin of the strongly variable $\gamma$-ray fluxes. These behaviors might also be reconciled in highly magnetized Poynting jet models,  mini-jet models, or turbulent cells \citep[e.g.,][]{gub09,mj10,nal11} that have been considered in TeV BL Lac objects. \citet{nal12} conclude that self-collimating jet structures might produce the conditions needed to explan the observations. 

Here we propose a new technique to produce rapid variability of VHE $\gamma$ rays far from the central black hole as a result of UHECR neutron production in the inner jet, with the outflowing neutrons making cascade synchrotron radiation from leptons formed as secondaries from photopion interactions of the UHECR neutrons with the dust IR radiation field at the pc scale. The power in escaping neutrons is accompanied by  $\gamma$-ray and neutrino radiations from the decay secondaries of neutral and charged pions. { This approach to variability, which depends on rectilinear propagation of the particles to large distances, has similarities with an idea of \citet{ghi09} where rapid variability of TeV blazars like PKS 2155-304 or Mkn 501 results from electrons forced to follow field lines. This magnetocentrifugal model would not likely work with 4C +21.35 at the pc scale, but also involves a geometry essentially different from a relativistic plasma. }

This model predicts neutrino fluxes detectable with IceCube from flares of FSRQs, not only PKS 1222+216, but also sources like 3C 279 and PKS 1510-089 that generate a VHE fluence reaching $\approx 10^{-3}$ erg cm$^{-2}$. For the one-half hour of activity of 4C +21.35, the total electromagnetic fluence is $\approx 10^{-5}$ erg cm$^{-2}$, but the flare itself could have lasted for a much longer  time than the half hour during which MAGIC was observing. Moreover, the long-term jet radiation could accelerate UHECRs in the inner jet with associated neutrino production from FSRQs with VHE emission.  Furthermore, UHECR protons and ions could escape upstream to produce emissions at the pc scale without significant $\gamma$-ray and neutrino production in the inner jet, but would be strongly depleted by interactions in the dust torus.

{ In conclusion, the proposed model, if correct,
would provide an important clue to the question of the origin of UHECRs. The feature of transport of inner jet energy to the pc scale via UHECRs would  explain why the question of the location of the $\gamma$-ray emission site in blazars has been so puzzling. UHECR processes could be essential in fitting the SED of all blazars, not just the FSRQs from which VHE radiation has been detected. }

\acknowledgements 
We would like to express our thanks to the referee, Dr. Boris Stern, for a valuable report, in particular, pointing out the importance of presenting the combined production spectra  for inner-jet physics from both BLR and IR torus target photon fields.  We  would also like to thank Armen Atoyan, Justin Finke, and Soebur Razzaque for discussions.  The work of C.D.D.\ is supported by the Office of Naval Research. K.M.\ is supported by JSPS and CCAPP. 

\appendix

\section{Derivation of the Photon Power}
\label{app:A}

The  apparent isotropic bolometric synchrotron luminosity is related to $L_{\rm syn}^\prime$,
 the isotropic synchrotron luminosity in the comoving jet frame,  according to the relation $L_{\rm syn}(\Omega)  = \delta_{\rm D}^4 L_{\rm syn}^\prime(\Omega^\prime) =  \delta_{\rm D}^4 (L_{\rm syn}^\prime/4\pi) = \delta_{\rm D}^4 L_{\rm syn}^\prime/4\pi$. The absolute synchrotron power radiated in all directions is, after multiplying by a factor of 2 for a two-sided jet, given by
\beq
L_{\rm syn, abs} = \oint d\Omega L_{\rm syn}(\Omega) = 2\times 2\pi \left({L_{\rm syn}^\prime\over 4\pi} \right)\,\int_{-1}^1 d\mu \,\delta_{\rm D}^4\;.
\label{Lsynabs}
\eeq
This can also be written as 
\beq
L_{\rm syn, abs} = {L_{\rm syn}^\prime\int_{-1}^1 d\mu \,\delta_{\rm D}^4\over L_{\rm syn}} \;L_{\rm syn}\;=\; 
{2\Gamma^2(3+\beta^2)\over 3\delta_{\rm D}^4} L_{\rm syn}\stackrel{\beta\rightarrow 1}{\rightarrow }{8\Gamma^2\over 3\delta_{\rm D}^4} L_{\rm syn}\;,
\label{Lsynabs1}
\eeq
noting $L_{syn} = 4\pi L_{syn}(\Omega )$. This beaming factor also applies to SSC emissions in the standard one-zone blazar model.

The  apparent isotropic bolometric Compton luminosity resulting from an isotropic comoving electron distribution scattering an external isotropic radiation field transforms according to $L_{\rm EC}\propto \delta_{\rm C}^6 L^\prime_{\rm EC}$.
Following the same reasoning, therefore, the absolute Compton power radiated in all directions is given in terms of the apparent Compton luminosity
by the relation
\beq
L_{\rm EC, abs} = { \int_{-1}^1 d\mu \,\delta_{\rm D}^6\over \delta_{\rm EC}^6} \;L_{\rm EC}\;=\; 
{2\Gamma^4(5+10\beta^2 + \beta^4)\over 5\delta_{\rm D}^6} L_{\rm EC}\stackrel{\beta\rightarrow 1}{\rightarrow }{32\Gamma^4\over 5\delta_{\rm D}^6} L_{\rm EC}\;.
\label{Lsynabs2}
\eeq
These relations apply to bolometric luminosities. Measurements over a finite frequency range introduce integration limits that give corrections that depend on spectral parameters.

\section{UHE Neutron Beam Production in Inner Jets}
\label{app:B}

UHECR acceleration may occur in inner jets via shock acceleration or possibly magnetic reconnection. 
For $t_{\rm var}=600$~s, the comoving size of the emission region is ${r'}_b = \dD c t_{\rm var}/(1+z) \simeq 3.6 \times {10}^{14} (\dD/20) ({t}_{\rm var}/600~{\rm s})/(1+z)$~cm. 
As long as we consider blob radii smaller than the radius of the BLR region, the photon field in inner jets is important.  Though the nonthermal photon field in inner jet is uncertain, following Tavecchio et al. (2010), we take the maximum allowed photon field as $L_{\rm syn}^b = {10}^{45.5}~{\rm erg}~{\rm s}^{-1}$ at $\nu_b ={10}^{14}$~Hz and a photon index $\beta=2.2$ at $\nu > \nu_b$ for demonstration.  The effective photon energy density at $\nu_b$ {in the inner jets is }
\begin{equation}
n_{eff} = \frac{u_{\rm syn}^b}{m_ec^2 \epsilon_b} \simeq 1.2 \times {10}^{13}~{\rm cm}^{-3} L_{\rm syn,45.5}^b {(\dD/20)}^{-5} \nu_{b,14}^{-1} {(t_{\rm var}/600~\rm s)}^{-2} {(1+z)}^2
\end{equation}
Then, the photomeson production efficiency in the inner jets is~\citep[e.g.,][]{ad03,mb10} 
\begin{eqnarray}
\eta_{p \gamma} &\approx& \frac{2 (u_{syn}^b/\epsilon_b)}{(1+\beta) m_e c^2} c \hat{\sigma}{\left( \frac{E_p}{E_p^b}\right)}^{1.2} \nonumber 
\\  
&\simeq& 0.2 {(\dD/20)}^{-4} \nu_{b,14}^{-1} {(t_{\rm var}/600~\rm s)}^{-1} L_{45.5}^b {(E/E_p^b)}^{1.2}  
\end{eqnarray}
where $E_p^b \simeq 1.5 \times {10}^{20}~{\rm eV}~{(\dD/20)}^2 \nu_{b,14}^{-1}$. This implies that $\eta_{p \gamma} \sim 8 \times {10}^{-3}$ at $E_p\sim{10}^{19}$~eV, which is not so efficient. 
Note that the corresponding photomeson production efficiency due to external photon fields (during the dissipation) would be low for smaller values of $\dD$.  In fact, for interactions with the dust photon field, we have 
\begin{equation}
\eta_{p \gamma} \sim 3 \times {10}^{-3} {(u_0/\epsilon_\star)}_4 (R/{10}^{15.5}~{\rm cm}),
\end{equation} 
where $R \approx \Gamma {r'}_b$ is the distance to the blob. 

If neutron production mainly occurs in such a compact region, we have the neutron conversion efficiency as
\begin{equation}
\eta_{p\gamma \rightarrow n} \approx 2.5 \eta_{p \gamma} \simeq 0.48  {(\dD/20)}^{-4} \nu_{b,14}^{-1} {(t_{\rm var}/600~\rm s)}^{-1} L_{45.5}^b {(E/E_p^b)}^{1.2}  
\end{equation} 
High-energy neutrons are depleted via conversion to protons or further photomeson production loss.  The neutron absorption efficiency $\eta_{n \gamma \rightarrow p}$ is expected to be comparable to $\eta_{p\gamma \rightarrow n}$, so the critical energy (where $\eta_{n \gamma \rightarrow p}=1$) is estimated to be  
\begin{equation} 
E_{n}^c \simeq 2.3 \times {10}^{20}~{\rm eV}~ {(\dD/20)}^{16/3} \nu_{b,14}^{-1/6} {(t_{\rm var}/600~\rm s)}^{5/6} {(L_{45.5}^b)}^{-5/6}   
\end{equation}
As a result, the resulting neutron spectrum can be written as 
\begin{equation}
E_n^2 N_n (E_n) \approx \frac{{\rm min}[1,\eta_{p \gamma \rightarrow n}]}{1+\eta_{n \gamma \rightarrow p}} E_p^2 N_p (E_p) \propto  E_p^{1-s+\beta},
\end{equation}
where we have assumed that the maximum proton energy is lower than $E_n^c$ for the last expression. The Hillas condition gives $E_n^{\rm max} \approx 0.8 E_p^{\rm max} \sim {10}^{18.5}$~eV, below which we expect $E_n^2 N_n \propto E^{1.2}$ for $s=2$. However, in our cases, the neutron beam production in inner jets typically requires quite large cosmic-ray luminosities.  On the other hand, one may also expect proton escape especially at the maximum energies. But details depend on the uncertain escape mechanism.  



\begin{thebibliography}{}
\bibitem[Abdo et al.(2011a)]{2011ApJ...733L..26A} Abdo, A.~A., Ackermann,  M., Ajello, M., et al.\ 2011a, \apjl, 733, L26
\bibitem[Abdo et al.(2011b)]{2011ApJ...736..131A} Abdo, A.~A., Ackermann, M., Ajello, M., et al.\ 2011b, \apj, 736, 131 
\bibitem[Abdo et al.(2010a)]{2010ApJ...710.1271A} Abdo, A.~A., Ackermann, M., Ajello, M., et al.\ 2010a, \apj, 710, 1271 
\bibitem[Abdo et al.(2010b)]{2010ApJ...721.1425A} Abdo, A.~A., Ackermann, 
M., Agudo, I., et al.\ 2010b, \apj, 721, 1425 \bibitem[Abdo et al.(2010c)]{2010ApJ...720..912A} Abdo, A.~A., Ackermann, 
M., Ajello, M., et al.\ 2010c, \apj, 720, 912 
\bibitem[Abdo et al.(2009)]{2009ApJ...699..817A} Abdo, A.~A., Ackermann, M., Ajello, M., et al.\ 2009, \apj, 699, 817 
\bibitem[Ackermann et al.(2011)]{2011ApJ...743..171A} Ackermann, M., Ajello, M., Allafort, A., et al.\ 2011, \apj, 743, 171  (2LAC)
\bibitem[Ackermann et al.(2010)]{2010ApJ...721.1383A} Ackermann, M., Ajello, M., Baldini, L., et al.\ 2010, \apj, 721, 1383 
\bibitem[Actis et al.(2011)]{cta10} Actis, M., et al. 2011, Exp. Astron., 32, 193
\bibitem[Aharonian et al.(2007)]{aha07} Aharonian, F., et al.\ 2007, \apjl, 664, L71 (PKS 2155-304)
\bibitem[Albert et al.(2007)]{alb07} Albert, J., et al.\ 2007, \apj, 669, 862 (Mrk 501)
\bibitem[Albert et al.(2008)]{alb08} Albert, J., et al.\ 2008, Science, 320, 1752; also arXiv:1101.2522
\bibitem[Aleksi{\'c} et al.(2011a)]{ale11a} Aleksi{\'c}, J., et al.\ 2011a, \aap, 530, A4 (3C 279)
\bibitem[Aleksi{\'c} et al.(2011b)]{ale11b} Aleksi{\'c}, J., et al.\ 2011b, \apjl, 730, L8 (PKS 1222+21)
\bibitem[Atoyan \& Aharonian(1996)]{aa96} Atoyan, A.~M., \& Aharonian, F.~A.\ 1996, \mnras, 278, 525 
\bibitem[Atoyan \& Dermer(2001)]{ad01} Atoyan, A., \& Dermer, C.~D.\ 2001, \prl, 87, 221102
\bibitem[Atoyan \& Dermer(2003)] {ad03} Atoyan, A.~M., \& Dermer, C.~D.\ 2003, \apj, 586, 79 
\bibitem[Atwood et al.(2009)]{atw09} Atwood, W.~B., et al.\ 2009, \apj, 697, 1071 
\bibitem[Begelman et al.(1990)]{brs90} Begelman, M.~C., Rudak, B., \& Sikora, M.\ 1990, \apj, 362, 38 
\bibitem[Begelman et al.(2008)]{bfr08} Begelman, M.~C., Fabian, A.~C., \& Rees, M.~J.\ 2008, \mnras, 384, L19 
\bibitem[Berezinsky et al.(2006)]{bgg06} Berezinsky, V., Gazizov, A., \& Grigorieva, S.\ 2006, \prd, 74, 043005 
\bibitem[B{\l}a{\.z}ejowski et al.(2000)]{2000ApJ...545..107B} 
B{\l}a{\.z}ejowski, M., Sikora, M., Moderski, R., \& Madejski, G.~M.\ 2000, \apj, 545, 107 

\bibitem[B\"ottcher(2010)]{2010arXiv1006.5048B} B\"ottcher, M.\ 2010, In ``Fermi Meets Jansky - AGN at Radio and Gamma-Rays", Eds.: Savolainen, T., Ros, E., Porcas, R. W., and Zensus, J. A., p. 41 (2010) arXiv:1006.5048  
\bibitem[B{\"o}ttcher et al.(2009)]{brs09} B{\"o}ttcher, M., 
Reimer, A., \& Marscher, A.~P.\ 2009, \apj, 703, 1168 
\bibitem[Bonnoli et al.(2011)]{bon11} Bonnoli, G., Ghisellini, G., Foschini, L., Tavecchio, F., 
\& Ghirlanda, G.\ 2011, \mnras, 410, 368 
\bibitem[Bromberg \& Levinson(2009)]{bl09} Bromberg, O., \& Levinson, A.\ 2009, \apj, 699, 1274 
\bibitem[Brown, Mikaelian, \& Gould(1973)]{bmg73} Brown, R.~W., Mikaelian, K.~O., \& Gould, R.~J.\ 1973, {\it Astrophysics Letters}, 14, 203 
\bibitem[Carrasco et al.(2010)]{car10}Carrasco, L., Carrami{\~n}ana, A., Recillas, E., Porras, A., \& Mayya, D.~Y.\ 2010, ATel, \#2626
\bibitem[Celotti 
\& Fabian(1993)]{1993MNRAS.264..228C} Celotti, A., \& Fabian, A.~C.\ 1993, \mnras, 264, 228 
\bibitem[Celotti 
\& Ghisellini(2008)]{2008MNRAS.385..283C} Celotti, A., \& Ghisellini, G.\ 2008, \mnras, 385, 283 
\bibitem[Cortina et al.(2012)]{cor12} Cortina, J., et al.\ 2012, ATel \#3965 (PKS 1510-089)
\bibitem[Dermer(1995)]{der95} Dermer, C.~D.\ 1995, \apjl, 446, L63 
\bibitem[Dermer \& Atoyan(2004)]{da04} Dermer, C.~D., \& Atoyan, A.\ 2004, \aap, 418, L5 
\bibitem[Dermer \& B\"ottcher(2006)]{db06} Dermer, C.~D., \& B\"ottcher, M.\ 2006, \apj, 643, 1081 
\bibitem[Dermer \& Menon(2009)]{dm09} Dermer, C.~D., \& Menon, G.\ 2009, High Energy Radiation from Black Holes: Gamma Rays, Cosmic Rays, and Neutrinos (Princeton Univerisity Press: Princeton)
\bibitem[Dermer et al.(2007)]{drl07} Dermer, C.~D., Ramirez-Ruiz, E., \& Le, T.\ 2007, \apjl, 664, L67 \bibitem[Dermer et al.(2009)]{der09} Dermer, C.~D., Razzaque, S., Finke, J.~D., \& Atoyan, A.\ 2009, New Journal of Physics, 11, 065016
\bibitem[Dermer \& Razzaque(2010)]{dr10} Dermer, C.~D., \& Razzaque, S.\ 2010, \apj, 724, 1366 
\bibitem[Dermer \& Schlickeiser(2002)]{ds02} Dermer, C.~D., \& Schlickeiser, R.\ 2002, \apj, 575, 667
\bibitem[Dom{\'{\i}}nguez et al.(2011)]{2011MNRAS.410.2556D} 
Dom{\'{\i}}nguez, A., Primack, J.~R., Rosario, D.~J., et al.\ 2011, \mnras, 
410, 2556 
\bibitem[Essey \& Kusenko(2010)]{ek10}Essey, W., \& Kusenko, A. 2010, Astropart. Phys., 33, 81
\bibitem[Essey et al.(2010)]{ess10}  Essey, W., Kalashev, O.~E., Kusenko, A., \& Beacom, J.~F.\ 2010, \prl, 104, 141102 
\bibitem[Essey et al.(2011)]{ess11} Essey, W., Kalashev, O., Kusenko, A., \& Beacom, J.~F.\ 2011, \apj, 731, 51 
\bibitem[Fan et al.(2006)]{2006ApJ...646....8F} Fan, Z., Cao, X., \& Gu, M.\ 2006, \apj, 646, 8 
\bibitem[Farrar \& Gruzinov(2009)]{fg09} Farrar, G.~R., \& Gruzinov, A.\ 2009, \apj, 693, 329 
\bibitem[Finke et al.(2008)]{fdb08} Finke, J.~D., Dermer, C.~D., B\"ottcher, M.\ 2008, \apj, 686, 181 
\bibitem[Finke \& Dermer(2010)]{fd10} Finke, J.~D., \& Dermer, C.~D.\ 2010, \apjl, 714, L303 
\bibitem[Fossati et al.(2008)]{fos08} Fossati, G., et al.\ 2008, \apj, 677, 906 (Mrk 421)
\bibitem[Foschini(2011)]{fos11} Foschini, L.\ 2011,  arXiv:1103.2008 
\bibitem[Foschini et al.(2011a)]{2011A&A...530A..77F} Foschini, L., Ghisellini, G., Tavecchio, F., Bonnoli, G., \& Stamerra, A.\ 2011a, \aap, 530, A77 
\bibitem[Foschini et al.(2011b)]{2011arXiv1110.4471F} Foschini, L., Ghisellini, G., Tavecchio, F., Bonnoli, G., \& Stamerra, A.\ 2011b, arXiv:1110.4471 
\bibitem[Francis et al.(1991)]{fra91} Francis, P.~J., Hewett, P.~C., Foltz, C.~B., et al.\ 1991, \apj, 373, 465
\bibitem[Gabici \& Aharonian(2005)]{ga05} Gabici, S., \& Aharonian, F.~A.\ 2005, \prl, 95, 251102 
\bibitem[Georganopoulos et al.(2001)]{gkm01} Georganopoulos, M., Kirk, J.~G., \& Mastichiadis, A.\ 2001, \apj, 561, 111 
\bibitem[Georganopoulos \& Kazanas(2003)]{gk03} Georganopoulos, M., \& Kazanas, D.\ 2003, \apjl, 594, L27 
\bibitem[Georganopoulos et al.(2004)]{gkm04} Georganopoulos, M., Kirk, J.~G., \& Mastichiadis, A.\ 2004, \apj, 604, 479 
\bibitem[Ghisellini et al.(2005)]{gtc05} Ghisellini, G., Tavecchio, F., \& Chiaberge, M.\ 2005, \aap, 432, 401 
\bibitem[Ghisellini \& Tavecchio(2008)]{gt08} Ghisellini, G., \& Tavecchio, F.\ 2008, \mnras, 387, 1669 
\bibitem[Ghisellini et al.(2009)]{ghi09} Ghisellini, G., 
Tavecchio, F., Bodo, G., \& Celotti, A.\ 2009, \mnras, 393, L16 
\bibitem[Giannios et al.(2009)]{gub09} Giannios, D., Uzdensky, D.~A., \& Begelman, M.~C.\ 2009, \mnras, 395, L29 
\bibitem[Gould \& Schr{\'e}der (1967)]{gs67}Gould, R.~J., \& Schr{\'e}der, G.~P.\ 1967, Phys.~ Rev., 155, 1404 
\bibitem[Gould(1979)]{1979A&A....76..306G} Gould, R.~J.\ 1979, \aap, 76, 306 
\bibitem[Hillas(1984)]{hil84} Hillas, A.~M.\ 1984, \araa, 22, 425 
\bibitem[Kneiske \& Dole(2010)]{kd10} Kneiske, T.~M., \& Dole, H.\ 2010, \aap, 515, A19 
\bibitem[Kotera et al.(2011)]{kal11} Kotera, K., Allard, D., \& Lemoine, M.\ 2011, \aap, 527, A54 
\bibitem[Lind 
\& Blandford(1985)]{lb85} Lind, K.~R., \& Blandford, R.~D.\ 1985, \apj, 295, 358 
\bibitem[Malmrose et al.(2011)]{mal11} Malmrose, M.~P., Marscher, A.~P., Jorstad, S.~G., Nikutta, R., \& Elitzur, M.\ 2011, \apj, 732, 116 
\bibitem[Mannheim \& Biermann(1992)]{mb92} Mannheim, K., \& Biermann, P.~L.\ 1992, \aap, 253, L21 
\bibitem[Mariotti et al.(2010)]{mar10} Mariotti, M., et al.\ ATel \#2684 
\bibitem[Marscher(2006)]{mar06} Marscher, A.~P.\ 2006, 
Relativistic Jets: The Common Physics of AGN, Microquasars, and Gamma-Ray 
Bursts, 856, 1 
\bibitem[Marscher 
\& Jorstad(2010)]{mj10} Marscher, A.~P., \& Jorstad, S.~G.\ 2010, arXiv:1005.5551 
\bibitem[M{\"u}cke et al.(2000)]{muc+00}
M{\"u}cke, A., Engel, R., Rachen, J. P., Protheroe, R. J., \& Stanev, T. 2000, Comput. Phys. Commun., 124, 290
\bibitem[M{\"u}cke \& Protheroe(2001)]{2001ICRC....3.1153M} M{\"u}cke, A., \& Protheroe, R.~J.\ 2001, International Cosmic Ray Conference, 3, 1153 
\bibitem[Murase(2007)]{mur07} Murase, K.\ 2007, \prd, 76, 123001 
\bibitem[Murase(2009)]{mur09} Murase, K.\ 2009, Physical Review Letters, 103, 081102 
\bibitem[Murase(2012)]{mur12} Murase, K.\ 2012, \apjl, 745, L16
\bibitem[Murase \& Beacom(2010)]{mb10} Murase, K., \& Beacom, J.~F. \ 2010, \prd, 81, 123001 
\bibitem[Murase \& Takami(2009)]{mt09} Murase, K., \& Takami, H.\ 2009, \apjl, 690, L14 
\bibitem[Murase et al.(2012)]{mdtm11} Murase, K., Dermer, C.~D., Takami, H., \& Migliori, G.\ 2012, \apj, 749, 63 



\bibitem[Nalewajko et al.(2011)]{nal11} 
Nalewajko, K., Giannios, D., Begelman, M.~C., Uzdensky, D.~A., \& Sikora, M.\ 2011, \mnras, 413, 333 
\bibitem[Nalewajko et al.(2012)]{nal12} Nalewajko, K., Begelman, M.~C., Cerutti, B., Uzdensky, D.~A., \& Sikora, M.\ 2012, arXiv:1202.2123 
\bibitem[Narayan \& Piran(2012)]{2012MNRAS.420..604N} Narayan, R., \& Piran, T.\ 2012, \mnras, 420, 604 

\bibitem[Neronov, Semikoz \& Vovk (2010)]{nsv10} Neronov, A., Semikoz, D., \& Vovk, Ie., ATEL \#2617
\bibitem[Poutanen \& Stern(2010)]{ps10} Poutanen, J., \& Stern, B.\ 2010, \apjl, 717, L118 
\bibitem[Razzaque et al.(2004)]{rmz04} Razzaque, S., M{\'e}sz{\'a}ros, P., \& Zhang, B.\ 2004, \apj, 613, 1072 
\bibitem[Shaw et al.(2012)]{2012ApJ...748...49S} Shaw, M.~S., Romani, R.~W., Cotter, G., et al.\ 2012, \apj, 748, 49 
\bibitem[Sikora et al.(2009)]{2009ApJ...704...38S} Sikora, M., Stawarz,  {\L}., Moderski, R., Nalewajko, K., \& Madejski, G.~M.\ 2009, \apj, 704, 38 
\bibitem[Stern \& Poutanen(2011)]{sp11} Stern, B.~E., \& Poutanen, J.\ 2011, \mnras, 417, L11 
\bibitem[Tanaka et al.(2011)]{tan11} Tanaka, Y.~T., et al.\ 2011, \apj, 733, 19 
\bibitem[Tavecchio et al.(2010)]{tav10}Tavecchio, F., Ghisellini, G., Bonnoli, G., \& Ghirlanda, G.\ 2010, \mnras, 405, L94
\bibitem[Tavecchio et al.(2011)]{tav11} Tavecchio, F., Becerra-Gonzales, J., Ghisellini, G., Stamerra, A., Bonnoli, G., Foschini, L., \& Maraschi, L.\ 2011, \aap, 534, A86 
\bibitem[Tavecchio et al.(2012)]{tav12} Tavecchio, F., 
Roncadelli, M., Galanti, G., \& Bonnoli, G.\ 2012, arXiv:1202.6529 

\bibitem[Wagner et al.(2010)]{wag10} Wagner, S.~J., Behera, B., \& HESS collaboration 2010, Bulletin of the American Astronomical Society, 42, 660 
\bibitem[Wang et al.(2004)]{2004ApJ...615L...9W} Wang, J.-M., Luo, B., \& Ho, L.~C.\ 2004, \apjl, 615, L9 
\bibitem[Waxman(2004)]{wax04} Waxman, E.\ 2004, New Journal 
of Physics, 6, 140 

\end{thebibliography}
\end{document}